\documentclass[authoryear,3p,twocolumn,10pt]{elsarticle}

\usepackage{lineno}

\usepackage{natbib}

\bibpunct{(}{)}{;}{a}{,}{,}

\usepackage{graphicx}
\usepackage{amssymb,array}
\usepackage{amsmath,amsthm,wasysym}

\begin{document}

\begin{frontmatter}



\title{Stability of Ice/Rock Mixtures with Application to a \\Partially Differentiated Titan}


\author{Joseph G. O'Rourke}
\author{David J. Stevenson}

\address{Division of Geological and Planetary Sciences, \\ 
California Institute of Technology, Pasadena, CA 91125, USA}

%% ----- ELSEVIER STUFF -----
%% The commands below up to the \end{frontmatter} are commented out
%% so that we can do some Icarus-required formatting on the second and
%% third pages that is not required later on by Elsevier.  So when
%% your paper gets accepted, and you are ready to start dealing with
%% Elsevier, copy your abstract and keywords up here, uncomment these
%% lines, and comment out the ICARUS STUFF below.
%% 
%% Alternately, you might just want to move these abstract, keyword,
%% and end frontmatter commands down, and comment out the ICARUS STUFF
%% commands.  It doesn't matter.

% \begin{abstract}
% % Text of abstract
% 
% \end{abstract}
% 
% \begin{keyword}
% % keywords here, in the form: keyword \sep keyword
% 
% 
% % PACS codes here, in the form: \PACS code \sep code
% 
% \end{keyword}

\begin{abstract}

Titan's moment of inertia, calculated assuming hydrostatic equilibrium from gravity field data obtained during the Cassini-Huygens mission, implies an internal mass distribution that may be incompatible with complete differentiation. This suggests that Titan may have a mixed ice/rock core, possibly consistent with slow accretion in a gas-starved disk, which may initially spare Titan from widespread ice melting and subsequent differentiation. A partially differentiated Titan, however, must still efficiently remove radiogenic heat over geologic time. We argue that compositional heterogeneity in the major Saturnian satellites indicates that Titan formed from planetesimals with disparate densities. The resulting compositional anomalies would quickly redistribute to form a vertical density gradient that would oppose thermal convection. We use elements of the theory of double-diffusive convection to create a parameterized model for the thermal evolution of ice/rock mixtures with a stabilizing compositional gradient. To account for large uncertainties in material properties and accretionary processes, we perform simulations for a wide range of initial conditions. Ultimately, for realistic density gradients, double-diffusive convection in the ice/rock interior can delay, but not prevent, ice melting and differentiation, even if a substantial fraction of potassium is leached from the rock component. Consequently, Titan is not partially differentiated.

\end{abstract}

% %% Keywords should appear after the abstract. 

%% ----- END ELSEVIER STUFF -----

\end{frontmatter}

%main text
\section{Introduction}

Titan is Saturn's largest satellite, the second largest in our Solar System. The surface of Titan features stable liquid methane/ethane lakes, a hydrocarbon precipitation cycle comparable to Earth's hydrology, and myriad additional geologic features such as dune fields and mountainous terrain \citep{Jaumann2009}. From an astrobiology perspective, Titan is interesting as a modern analogue for pre-biotic Earth \citep{Raulin2009}. Titan has a thick nitrogen atmosphere, which must be sustained by the continuous production of methane \citep{Atreya2006}. Methane may be stored subsurface or perhaps more deeply in stable clathrate-hydrates along with other volatiles and episodically outgassed when internal heating causes melting \citep{Gautier2005,Tobie2006}. Understanding Titan's fascinating surface, atmosphere, and chemistry requires knowledge of the evolution of its interior. 

Measurements of the gravity field of Titan can place indirect constraints on the current structure of its interior. Titan's gravity harmonics were determined to degree 3 through careful tracking of the Cassini spacecraft during four flybys \citep{Iess2010}. The calculated ratio $J_2/C_{22}$ is consistent with the value of 10/3 that is itself consistent with a gravity field dominated by a nearly hydrostatic quadrupole, although non-zero values of other degree 2 and 3 coefficients indicate that nonhydrostatic features are present. If hydrostatic equilibrium is assumed, however, then Titan's moment of inertia (MoI) factor is found to be C $\sim$ 0.34. Here, the MoI is $CM_sR_s^2$, where $M_s$ and $R_s$ respectively represent the mass and radius of Titan. Measurement of the tidal Love number $k_2$ from additional flybys reveals a relatively large response of the gravity field to the Saturnian tidal field, indicating the presence of a global, subsurface ocean \citep{Iess2012}. The decoupling of Titan's shell from its interior with an ocean may also explain Titan's long-wavelength topography \citep{Nimmo2010}, spin pole orientation \citep{Bills2011}, and Schumann resonance \citep{Simoes2012}.

The 2-$\sigma$ error on $J_2/C_{22}$ is about 3$\%$, centered on 10/3 and dominated by the error in $J_2$, according to \citet{Iess2010}. We can write $J_2=J_{2,h}+J_{2,nh}$ and likewise $C_{22}=C_{22,h}+C_{22,nh}$, where ``h" means hydrostatic and ``nh" means non-hydrostatic. So, $J_{2,h}/C_{22,h}$ is \textit{exactly} 10/3. If the non-hydrostatic parts were completely uncorrelated (i.e., if they did not cancel in the ratio $J_2/C_{22}$), then the closeness of the observed value to 10/3 is highly significant and precludes a substantial error in MoI. If, on the other hand, the non-hydrostatic parts tend to be correlated (as they would, for example, in the tidal heating model of \citet{Nimmo2010} or if Titan has undergone True Polar Wander to the preferred orientation of the non-hydrostatic part of the MoI), then the small deviation away from 10/3 does not guarantee smallness of the non-hydrostatic part and the MoI is accordingly uncertain. The effects of the non-hydrostatic components on Titan's MoI are explored in detail in \citet{Gao2012}. Given the observed power in degree 3 gravity, \citet{Iess2010} propose that the MoI could be as low as 0.33 and we adopt this as a plausible lower bound. The true MoI is likely to be smaller than 0.34, the value inferred from $J_2/C_{22}$, because this is a lower rotational energy state. Since a fully differentiated Titan may allow 0.33 (but perhaps not 0.34), the differentiation state of Titan is accordingly not yet determined from observation.

Titan's MoI can be compared to previous results regarding other icy satellites. In particular, Titan's MoI coefficient is intermediate to the previously-measured C $\sim$ 0.31 for Ganymede \citep{Anderson1996} and C $\sim$ 0.35 for Callisto \citep{Anderson2001}, where hydrostatic equilibrium was assumed \textit{a priori} for the Galilean satellites. While Ganymede is easily modeled as a differentiated satellite with an iron core under a rock shell and an outer ice layer \citep[e.g.,][]{Sohl2002}, models of a differentiated Callisto are not consistent with the reported MoI \citep[e.g.,][]{Anderson2001}. A proposed interior structure for Callisto features a rock/ice core with rock mass fraction near the close packing limit, with an overlying icy mantle that was depleted of rock by Stokes settling \citep{Nagel2004}. In this case, an ice/rock lithosphere in which density decreases with depth overlays the icy mantle. 

Titan might represent an intermediate case between differentiated Ganymede and undifferentiated Callisto. However, the inferred (and widely cited) partially differentiated state for Callisto is based on a gravity inversion that \textit{assumes} $J_2/C_{22}$ = 10/3, and perhaps the inferred MoI is incorrect. As for Titan, the sense of the error is likely to cause an overestimate of the MoI because non-hydrostatic contributions to $J_2$ and $C_{22}$ are likely positive. At present, we cannot exclude the possibility that \textit{all} large icy satellites are fully differentiated. (In this paper, we use that term to imply the complete separation of ice from rock; the further differentiation of an iron core from the rock is a separate issue that we do not address.) It should be noted that non-hydrostatic effects are more likely to be important in slowly rotating bodies (Titan and Callisto) relative to Ganymede because the hydrostatic effects scale as rotational frequency squared for both tides and synchronous rotation \citep{Gao2012}. Certainly, Callisto and Ganymede are different in appearance and Callisto, like Titan, lacks a global magnetic field. 

Relative differences in the experienced intensity of the Late Heavy Bombardment (LHB) may explain the Ganymede/Callisto dichotomy. In the so-called ``Nice model," the gas giant planets swiftly realigned once Jupiter and Saturn crossed their mean motion resonance $\sim$700~Myr after planetary formation, quickly causing the outward migration of Uranus and Neptune and the evolution of Jupiter and Saturn's orbital eccentricities \citep{Tsiganis2005}. The ensuing destabilization of the planetesimal disk and the asteroid belt caused the LHB of both the outer and inner solar system \citep{Gomes2005}. Ganymede is closer to Jupiter than Callisto, so it likely suffered considerably more high-energy impacts during the LHB. For assumptions about the planetesimal disk consistent with the Nice model, the differences in received energy during the LHB are sufficient to cause Ganymede to differentiate, while Callisto's rock and ice components may remain unmixed \citep{BarrCanup2010}. If Titan survived both accretion and the LHB without melting the ice in its deep interior, then it could have remained partially differentiated like Callisto, at least initially.

Gas giant satellites like Titan were accreted from the outskirts of the disk of material surrounding their parent planets. Nascent gas giant planets must accrete enough rock such that they can also accumulate large amounts of gas, principally hydrogen and helium, before the dissipation of the protoplanetary disk, which had a lifetime of a few Myr \citep[e.g.,][]{Lissauer2007,Lunine2010}. In the core-nucleated gas accretion model, for instance, the formation of Saturn began with the accretion of km-sized rocky planetesimals from the minimum mass sub nebula (MMSN) \citep[e.g.,][]{Lissauer2007}. However, if gas giant satellites accreted from the outskirts of a disk of planet-forming material as dense and gas-rich as the MMSN, then Jupiter's Galilean satellites (and, by analogy, Titan) would have formed quickly and hot and therefore differentiated. Additionally, recent work suggests that collisional mergers of Galilean-like satellites may have formed Titan \citep{Asphaug2013}, which would have caused complete differentiation. 

Internal structure models for Titan often assume widespread ice melting and thus differentiation. Many studies investigated the thermal evolution of Titan assuming a large silicate/iron core \citep[e.g.,][]{Sohl2003,Tobie2005}, before MoI data from \citet{Iess2010} cast doubt on such models. Papers invoked convection in a silicate core to melt clathrate hydrates and cause episodic outgassing of methane \citep[e.g.,][]{Tobie2006}. Another class of models assumes that Titan contains a large core of hydrated silicates, chiefly the serpentine mineral antigorite \citep{Fortes2007,Grindrod2008,Castillo2010,Fortes2012,Tobie2012}. Serpentinization of silicates in icy satellites is likely to be rapid in the presence of liquid water, e.g. during differentiation or subsequent hydrothermal convection \citep{Ransford1981,Travis2005}, because the serpentinization reaction promotes crack formation and increasing material permeability \citep{MacDonald1985}. An interior dominated by hydrated silicates is consistent with constraints on Titan's chemical evolution \citep[e.g.,][]{Fortes2007,Fortes2012}. But a partially differentiated Titan, in which Titan's deep interior is a mixture of ice and rock, is still consistent with the gravity data \citep{Iess2010} and such models have not been fully vetted.

The accretion of undifferentiated icy satellites is often assumed to occur in a ``gas-starved" disk, where material similar to that found in the gas giant's feeding zone is fed to the accretion process over millions of years. The resultant, long timescales for satellite accretion allow many of the Galilean satellites, particularly Callisto, to avoid complete differentiation \citep[e.g.,][]{Canup2002}. In another model, the major satellites of Jupiter and Saturn formed in a solid-enhanced minimum mass planetary nebula, avoiding differentiation as satellites opened gaps in the nebula \citep{Mosqueira2003a,Mosqueira2003b}. With a particular compositional gradient in the initial circumplanetary disc, the compositions of the major Saturnian satellites can be roughly reproduced \citep{Mosqueira2010}. In any model, accretion must also be delayed to escape intense, but short-lived, radiogenic heating from the decay of $^{26}$Al and $^{60}$Fe if an icy satellite is to avoid complete differentiation \citep[e.g.,][]{McKinnon1997,Barr2008}. Despite myriad threats to unmelted ice, slow accretion in a gas-starved disk permits the formation of a partially differentiated Titan \citep{Barr2010}.

The purpose of this study is to determine whether a partially differentiated Titan is stable over geologic time. After accretion, the internal ice/rock mixture must efficiently expel radiogenic heat, which would otherwise melt the ice and allow irreversible sinking and separation of the rock component. Because the major Saturnian satellites are remarkably heterogenous, we argue that Titan likely accreted from planetesimals with disparate rock mass fractions. Even without that heterogeneity, there may be a tendency to form a stably stratified interior at the outset. Soon after accretion, a stabilizing density gradient would be established in which rock mass fraction increases with depth. Aspects of the theory of double-diffusive convection allow us to formulate a one-dimensional, parameterized thermal evolution model in which a convecting layer slowly grows to encompass the entire ice/rock interior. We perform simulations for a large set of initial conditions because of uncertainties in key parameters and to analyze the sensitivity of our results. Finally, we discuss alternatives to a partially differentiated Titan and consider possible implications for other icy satellites.

\section{Theoretical formulation}

Our simple model assumes that Titan is a sphere of radius $R_s$ and mass $M_s$, consisting of an undifferentiated ice/rock interior with radius $R_i$ and an overlying shell. The structure of the outer shell may be complicated, especially considering the likely presence of an ocean, but we do not need to model it in detail. We perform parametrized thermal evolution simulations to assess whether Titan can remain partially differentiated until the present. Stabilizing compositional gradients, caused by increasing rock mass fraction and thus density with depth, likely inhibit convection in Titan's ice/rock interior. Elements from the theory of double-diffusive convection are incorporated to model thermal evolution with this complication. Although this simple model will not precisely reconstruct the thermal history of Titan, it does provide a first-order test of the hypothesis of partial differentiation.

\subsection{Establishment of vertical density gradients}

The major Saturnian satellites exhibit significant compositional heterogeneity. The average densities of each satellite vary widely, from $\sim$990~kg~m$^{-3}$ for Tethys to $\sim$1,600~kg~m$^{-3}$ for Enceladus \citep{Jacobson2004}. We can calculate a mass-weighted density:
\begin{linenomath*}
\begin{equation}
\bar\rho_{mw} = \frac{\sum_i m_i\bar\rho_i}{\sum_i m_i},
\end{equation}
\end{linenomath*}
where $m_i$ and $\bar\rho_i$ are the masses and mean densities of each satellite, respectively. Considering the eight major satellites excluding Titan (Mimas, Enceladus, Tethys, Dione, Rhea, Hyperion, Iapetus, and Phoebe), $\bar\rho_{mw}$ $\approx$ 1,270~kg~m$^{-3}$ \citep{Jacobson2004}. Titan's mean density is $\sim$1,881~kg~m$^{-3}$, but its uncompressed density is not much higher than the densities of the other Saturnian satellites. In fact, the inner satellites (the first five of the aforementioned) and Saturn's almost-pure water ice rings may have been formed from tidal stripping of a Titan-sized satellite that was lost within Saturn's classical Roche limit \citep{Canup2010}.

Since Titan was plausibly formed from planetesimals of size and composition similar to these satellites, lateral density anomalies as large as 10$\%$ likely existed after accretion. These unstable anomalies drive a flow that will restructure the ice/rock mixtures, so that more dense mixtures underlie less dense mixtures, in a length of time that scales as
\begin{linenomath*}
\begin{equation}
t_f \sim \frac{\mu}{\Delta\rho_{lat}gD}, 
\end{equation}
where $\mu$ is viscosity, $\Delta\rho_{lat}$ is the horizontal density anomaly, $g$ is gravitational acceleration, and $D$ is the length scale of the anomalies. A competing process is the Stokes settling of the dense rock fragments from the ice/rock mixture. Using the usual Stokes settling velocity, the time in which the rock fragments settle out of the mixture scales as 
\begin{equation}
t_s \sim \frac{\mu D}{\Delta\rho gR^2},
\end{equation} 
\end{linenomath*}
where $\Delta\rho$ is the density difference between the ice and the rock ($\sim$1,000 to 2,000~kg~m$^{-3}$) and $R$ is the radius of the rock fragments. For R $\sim$ 1~m and D $\sim$ 100~km, the formation of the vertical density gradients overwhelms the Stokes settling of the rocks if the lateral density anomaly is greater than only one part in ten billion of the density difference between the ice and rock, i.e. $t_f < t_s$ if $\Delta\rho_{lat}>10^{-10}\Delta\rho$. Clearly, this is easily achieved.

Mixing during accretion will not homogenize the differences arising from large planetesimals. Importantly, \textit{lateral} differences during the accretion lead to \textit{vertical} differences in ice/rock ratio after settling, even if individual impact events mix vertically. Therefore, models of Titan's thermal evolution must consider the existence of up to $\sim$10$\%$ stabilizing density gradients in Titan's ice/rock interior initially (and even smaller effects are still very important because thermal expansion of ice produces much smaller density differences). A vertical density gradient, where density increases with depth, opposes thermal buoyancy, inhibiting convection and thus militating against substantial heat transfer. If radiogenic heat is not removed from the system, then the ice in Titan's deep interior will melt, especially if low melting point components such as ammonia are present. The liberated rock components will sink as the melt percolates upwards. Runaway differentiation can occur if the gravitational energy released by the sinking of the rock component is sufficient to melt more ice \citep{Friedson1983}, but complete differentiation may occur without invoking runaway differentiation if convection is sufficiently suppressed.

\subsection{Double-diffusive convection applied to Titan}

The theory of double-diffusive convection was originally developed to elucidate various phenomena in oceanography \citep{Stern1960}. In general, double-diffusive phenomena occur in the presence of two opposing density gradients when the two responsible components have different molecular diffusivities \citep[e.g.,][]{Turner1974}. The theory has been extended to the dynamics of magma chambers and stellar interiors \citep[e.g.,][]{Spiegel1972,Schmitt1983,Turner1985}, but remains most often utilized in oceanography. For instance, in Earth's high latitudes, the polar oceans are composed of cold, fresh water overlying warm, salty water \citep[e.g.,][]{Schmitt1994}. In the presence of the stabilizing salinity gradient, convective motions resembling Rayleigh-B\'{e}nard convection arise after a stage of steady oscillating motions in layers cooled from above \citep{Veronis1968,Noguchi2010}. These well-mixed, convecting layers increase in thickness and frequently merge with neighboring layers of equal density. This is known as the ``diffusive-layer" mode, in which salinity, the stabilizing component, has a much lower molecular diffusivity than temperature, the component driving convection.

Titan's deep interior features analogous opposing temperature and compositional gradients. The compositional gradient is simply a measure of the increasing rock mass fraction in the ice/rock mixture with depth. Although a temperature gradient can induce particle motion through interfacial premelting \citep{Rempel2001}, the relevant time scales are far too long for application to icy satellites. Hence, the molecular diffusivity associated with the rock component is effectively zero. Because the molecular diffusivity of salt is non-zero, in contrast, complicated temperature and salinity profiles exist in the ocean \citep[e.g.,][]{Huppert1981}, with time-dependent fluxes of salt and temperature across diffusive interfaces between the multiple convecting layers \citep[e.g.,][]{Worster2004}. Regardless, the term ``double-diffusive" is still applicable to our model of convection in the presence of a stabilizing gradient. As outlined below, aspects of the theory applied to Titan do not depend on the molecular diffusivity of the component opposing temperature.

In Titan, double-diffusive convection occurs in the presence of a stabilizing rock mass fraction gradient as its deep interior is cooled from above and heated from within. The density within the ice/rock interior is a linear function of rock mass fraction and temperature \citep{Turner1974}:
\begin{linenomath*}
\begin{equation}
\rho = \rho_0(1-\alpha\Delta T+\beta\Delta S), \label{eq:rho}
\end{equation}
where $\rho_0$ is the density at some reference depth, $\alpha$ is the coefficient of thermal expansion, and $\beta = (1/\rho)(\partial\rho/\partial S)_T$ is the coefficient of compositional expansion, analogous to $\alpha$, associated with the change in rock mass fraction. The change of density due to pressure alone can be omitted because it does not contribute to the convective stability and is in any event too small to greatly affect the form of the convective motions. Table~\ref{table:const_Titan} contains numerical values of these coefficients and other constants used in this study. The rock mass fraction may be calculated \citep{Barr2010}: 
\begin{equation}
S=\frac{\rho_r(\bar\rho-\rho_i)}{\bar\rho(\rho_r-\rho_i)}, \label{eq:S}
\end{equation}
\end{linenomath*}
where $\bar\rho$ is the average density, $\rho_r$ is the density of rock, and $\rho_i$ is the density of ice. High-pressure ice phases V, VI, and VII may exist within Titan's deep interior; a representative density $\rho_i$ = 1,400~kg~m$^{-3}$ is chosen in accord with \citet{Barr2010}. Although the mineralogy of Titan's rock component is not completely known, CI carbonaceous chondrites and Prinn-Fegley rock, an assemblage that models condensate from the proto-Jovian nebula, are possible analogues, motivating the choice $\rho_r$ = 3,300~kg~m$^{-3}$ \citep{Mueller1988,Barr2010}.

\begin{table*} \centering
\begin{tabular}{{c}*{3}{l}*{1}{c}}
\hline
Constant & Definition & Value & Units & Ref. \\ \hline
$A$ & Activation parameter for ice & 25 & - & [1] \\
$\kappa$ & Thermal diffusivity & 10$^{-6}$ & m$^2$ s$^{-1}$ & [1] \\
$\alpha$ & Coefficient of thermal expansivity & 10$^{-4}$ & K$^{-1}$ & [1] \\
$\beta$ & Coefficient of compositional expansivity & 1.1 & - & $^*$\\
$M_s$ & Mass of Titan & 1.345 $\times$ 10$^{23}$ & kg & [2] \\
$R_s$ & Radius of Titan & 2575 & km & [2] \\
$R_i$ & Radius of ice/rock interior & 2000 & km & $^*$ \\
$\rho_i$ & Density of ice & 1400 & kg m$^{-3}$ & [2] \\
$\rho_r$ & Density of rock & 3300 & kg m$^{-3}$ & [2] \\
$C_{p,i}$ & Specific heat of ice & 2100 & J kg$^{-1}$ K$^{-1}$ & [2] \\
$C_{p,r}$ & Specific heat of rock & 900 & J kg$^{-1}$ K$^{-1}$ & [3] \\
\hline
\end{tabular}
\caption{List of key model constants. References: 1. \citet{Friedson1983}, 2. \citet{Barr2010}, 3. \citet{Grindrod2008}, $^*$this study.} \label{table:const_Titan} 
\end{table*}

A quick calculation with Eqs.~\ref{eq:rho} and~\ref{eq:S} reveals why compositional gradients make melting much more likely in a partially differentiated Titan. A $\sim$10$\%$ density contrast across Titan's interior corresponds to $\Delta S$ $\sim$ 10$^{-1}$ between ice/rock parcels at the top and bottom of the mixture. So, an (impossible) temperature increase of $\Delta T$ $\sim$ 10$^{3}$~K is required before compositionally dense parcels can become buoyant.

For cooling from above and heating from within, but with the top temperature kept constant, convection begins at the top of Titan's ice/rock interior. The onset of convection must occur at an upper thermal boundary layer because the temperature gradient throughout the ice/rock interior is initially adiabatic. With a linear density distribution, an equation for a measure of the initial stability gradient may be written \citep{Turner1973}:
\begin{linenomath*}
\begin{equation}
N_S^2 = -g\beta\frac{dS}{dz},
\end{equation}
where $N_S$ is a frequency of oscillation. When heat flows out of the top of the convecting layer, the associated buoyancy flux may be calculated \citep{Turner1973}:
\begin{equation}
B = -\frac{g\alpha H}{\rho C_p}.
\end{equation}
As cooling continues from above, a well-mixed convecting layer grows with height \citep{Turner1973}:
\begin{equation}
h = \left(\frac{2Bt}{N_S^2}\right)^{1/2}.
\end{equation}
Therefore, the rate of the growth of the convecting layer may be calculated:
\begin{equation}
\frac{dh}{dt}= \left[\frac{\alpha H}{2\rho C_p \beta (dS/dz)t}\right]^{1/2}.
\end{equation}
\end{linenomath*}

\begin{figure}
\noindent\includegraphics[width=20pc]{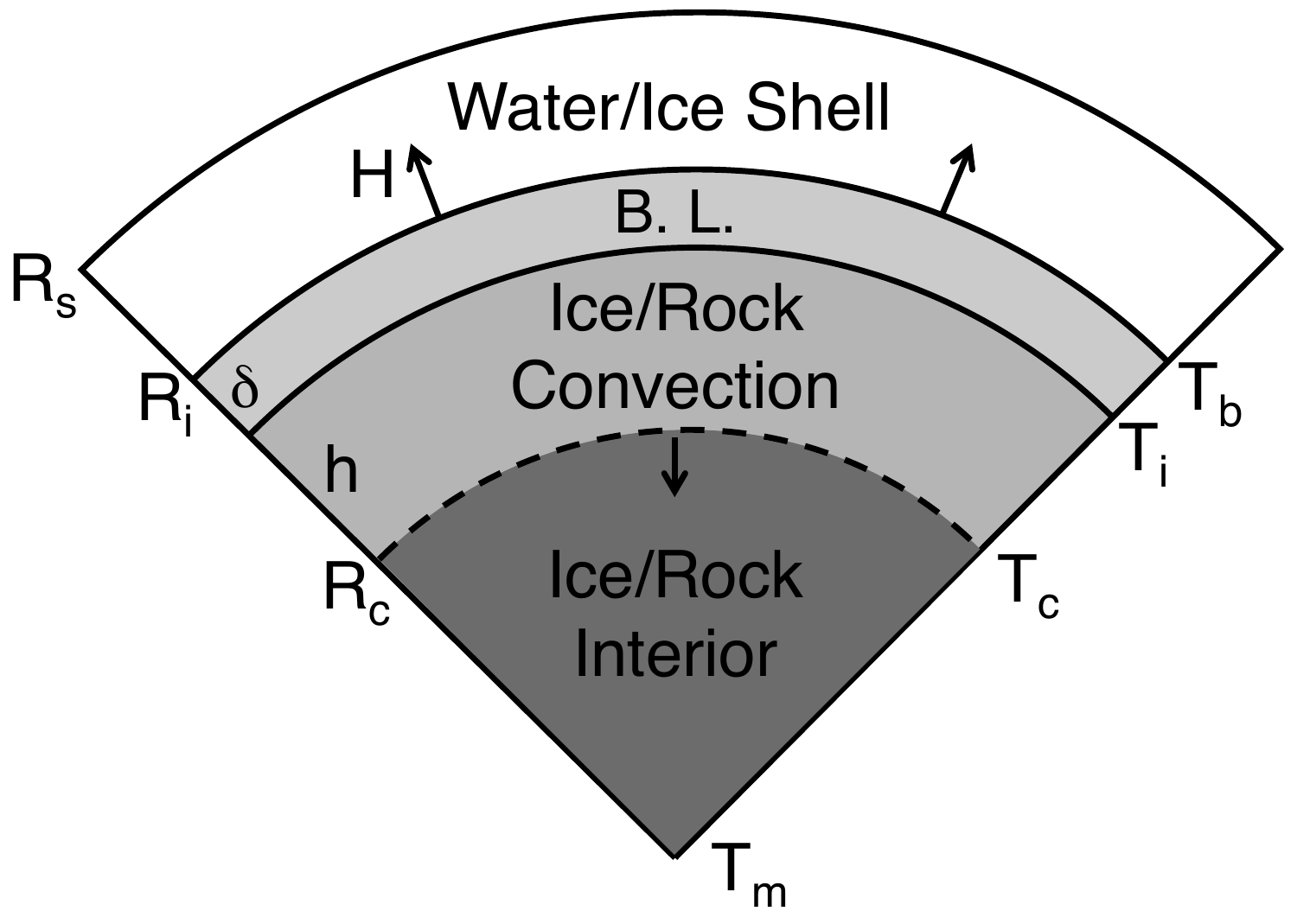}
\caption{Cartoon showing the assumed structure of Titan after the initialization of double-diffusive convection in the interior ice/rock region, before the convecting layer has grown to encompass the entire interior. Dimensions are not to scale and the overlying water/ice shell is not drawn in detail. Key model parameters are indicated as used in the text.} \label{fig:cartoon}
\end{figure}

Figure~\ref{fig:cartoon} shows the internal structure of Titan after double-diffusive convection has proceeded for some time. The radius of the ice/rock interior is $R_i$. The non-convecting ice/rock core has radius $R_c$, where $T_c$ and $T_m$ respectively represent the outer and inner temperatures. A thermal boundary layer exists between the convecting outer shell and the convecting ice/rock interior with thickness $\delta$. The temperature at the top of the thermal boundary layer, $T_b$, is held constant, while the temperature of the convecting layer is $T_i$. Parametrized thermal evolution models typically assume adiabatic temperature gradients in convecting layers, but these temperature increases with depth are negligible.

\subsection{Governing equations}

Energy in the convecting ice/rock layer is conserved:
\begin{linenomath*}
\begin{equation}
4\pi R^2_i H = \frac{4\pi}{3}\bar\rho (R_i^3 - R_c^3) \left(Q - C_p \frac{dT_i}{dt}\right)-\rho_if_iL_m, \label{eq:gov}
\end{equation}
where $H$ is the heat flux out of the convecting ice/rock layer into the overlying shell, $Q$ is radiogenic heat production, $f_i$ is the volume of melted ice, and $L_m$ is the latent heat of melting. The specific heat of the ice/rock mixture is a mass-weighted average \citep{Barr2008}:
\begin{equation}
C_p = SC_{p,r} + (1-S)C_{p,i},
\end{equation}
where $C_{p,r}$ and $C_{p,i}$ are the specific heats of the rock and ice components, respectively. Assuming that no melting occurs, we can conveniently rewrite Eq.~\ref{eq:gov}:
\begin{equation}
C_p\frac{dT_i}{dt}=Q - \frac{3H}{\bar\rho}\frac{R_i^2}{R_i^3-R_c^3}.
\end{equation}
\end{linenomath*}

We consider contributions to radiogenic heat production from $^{40}$K, $^{235}$U, $^{238}$U, and $^{232}$Th. Radiogenic heating in the ice/rock mixture may be calculated \citep[e.g.,][]{Barr2008}:
\begin{linenomath*}
\begin{equation}
Q(t) = S\sum_n c_{n,0}P_{n,0}\exp(-\lambda_nt), \label{eq:Q}
\end{equation}
\end{linenomath*}
where, for the \textit{n}th isotope, $c_{n,0}$ is the initial abundance, $P_{n,0}$ is the initial specific heat production, and $\lambda_n$ is the decay constant. Table~\ref{table:const_radio} lists the constants used to calculate radiogenic heat production. 

\begin{table}
\begin{tabular}{*{1}{l}*{3}{c}}
\hline
Isotope &  $c_{n,0}$ [ppb]& $P_{n,0}$ [$\mu$W/kg] & $\lambda_n$ [1/Gyr] \\ \hline
$^{238}$U & 16.91 & 93.7 & 0.155 \\
$^{235}$U & 5.275 & 569 & 0.985 \\
$^{232}$Th & 38.71 & 26.9 & 0.0495 \\
$^{40}$K $\rightarrow$ $^{40}$Ar & 813.8 & 1.02 & 0.0581 \\
$^{40}$K $\rightarrow$ $^{40}$Ca & 813.8 & 26.69 & 0.4962 \\ \hline
\end{tabular}
\caption{Constants used to calculate radiogenic heat production, adapted from \citet{Barr2008}.} \label{table:const_radio} \label{lasttable}
\end{table}

With this formulation, $t=0$~Gyr in Eq.~\ref{eq:Q} corresponds to the formation of the calcium-aluminum-rich inclusions (CAIs) in chondrites. If Titan formed exactly coincident with the CAIs, however, intense heating from $^{26}$Al and $^{60}$Fe would have caused widespread melting and hence differentiation \citep[e.g.,][]{Barr2010}. We implicitly assume that, in reality, Titan accreted several Myr after the formation of the CAIs and thus we may ignore these short-lived isotopes. A delay of a few Myr barely changes the amount of radiogenic heating from $^{40}$K, $^{235}$U, $^{238}$U, and $^{232}$Th. Because we do not aim to precisely constrain the timing of Titan's accretion, we simply repeat our simulations starting at $t_0$ = 5 and 10~Myr to confirm that our results do not change appreciably. Note that heating from $^{26}$Al and $^{60}$Fe would only lower the chances of survival for a partially differentiated Titan.

\begin{figure}
\noindent\includegraphics[width=20pc]{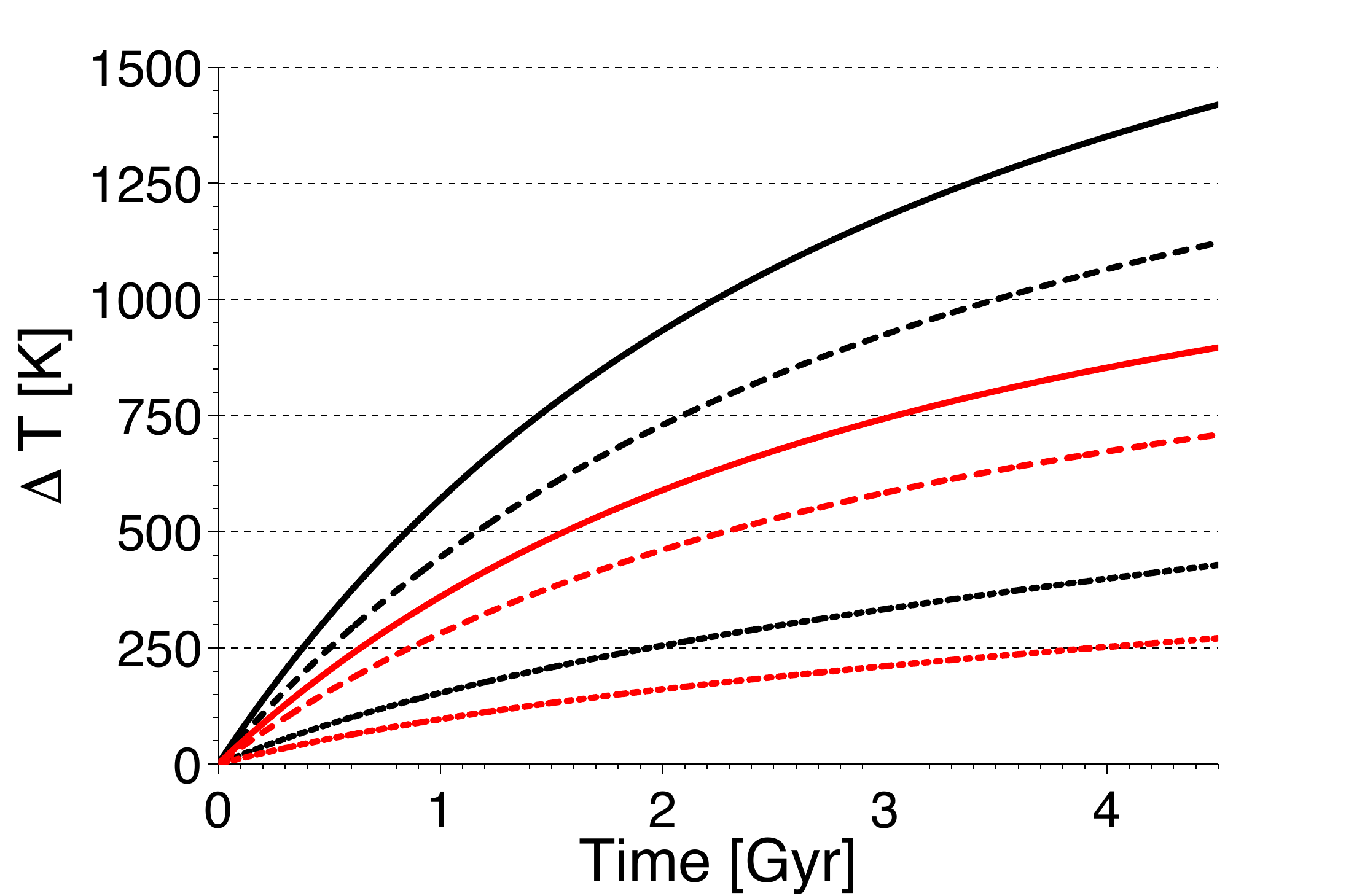}
\caption{Temperature increase in a parcel of rock (black, $S$ = 1.0) and in an ice/rock mixture with $S$ = 0.8 (red). Solid, dashed, and dotted lines respectively represent $f$ = 1.0, 0.7, and 0.3.} \label{fig:Q}
\end{figure}

Because contact between silicates and ammonia/water liquid can leach volatile $^{40}$K out of the rock component \citep[e.g.,][]{Engel1994}, we multiply the initial abundance of potassium by $f$, a depletion factor. Figure~\ref{fig:Q} shows representative temperature increases over 4.5~Gyr for $S$ = 1.0 and 0.8 from radiogenic heat production with $f$ = 1.0, 0.7, and~0.3.

\subsection{Thermal evolution of Titan's ice/rock interior}

A one-dimensional parametrized model can simulate the thermal evolution of the ice/rock mixture within Titan. The viscosity of the ice/rock mixture is a critical parameter, subject to large uncertainty. Although the deformation of ice is non-Newtonian under the conditions of Titan's deep interior, an Arrhenius-like equation may suffice for low strain rates \citep{Friedson1983}:
\begin{linenomath*}
\begin{equation}
\eta = \frac{\nu_0}{\rho}\exp\left[A\left(\frac{T_m}{T} - 1\right)\right], 
\end{equation}
\end{linenomath*}
where $A$ is an activation parameter and $T_m$ is the melting point of water ice, calculated using data from \citet{Petrenko1999}. For conditions appropriate to icy satellites, adding rock particles to a water ice matrix may cause viscosity to increase by an order of magnitude \citep{Friedson1983}, but we do not explicitly model these effects in this study.

At first, a conductive boundary layer grows with thickness $\delta \sim \sqrt{\kappa t}$ at the top of the ice/rock interior. The conductive heat flux out of the boundary layer is
\begin{linenomath*}
\begin{equation}
H = \frac{k \Delta T}{\delta},
\end{equation}
where $k=\kappa \rho C_p$ is thermal conductivity and $\kappa$ is thermal diffusivity.
The initial growth of the conductive boundary layer stops and convection begins when the Rayleigh number for the boundary layer reaches a critical value, i.e., 
\begin{equation}
Ra_c = \frac{g\alpha\Delta T \delta^3_c}{\eta(T) \kappa} = 10^3,
\end{equation}
\end{linenomath*}
where $\Delta T$ is the temperature contrast across the boundary layer and $\delta_c$ is the critical thickness of the boundary layer.

After convection begins, the viscosity contrast between the convecting ice/rock mixture and the overlying shell of high pressure ice is less than the $\sim$4 orders of magnitude required to produce a stagnant lid. So, the heat flux is given by the following equation \citep{Friedson1983}:
\begin{linenomath*}
\begin{equation}
H = 0.1 k (\Delta T)^{4/3}\left[\frac{g\alpha}{\eta(T_i)\kappa}\right]^{1/3}, \label{eq:H}
\end{equation}
where $\Delta T = T_b-T_i$ is the temperature contrast driving convection. The thickness of the thermal boundary layer, $\delta$, can be calculated using the condition for convective instability 
\begin{equation}
\frac{g\alpha\Delta T \delta^3_l}{\eta(T_{i}-0.5\Delta T) \kappa} = 10^3.
\end{equation}
\end{linenomath*}

In reality, the thermal evolution of the ice shell above the thermal boundary layer is complicated. Titan likely has a shell consisting of an outer ice I layer, an ocean, and a mantle of high-pressure ice polymorphs \citep[e.g.,][]{Iess2012}. Convection in the ice~I shell is expected for small grain sizes \citep{Barr2007,Mitri2008}, but the long-wavelength topography of Titan implies shell thickness variations that convection might eliminate, indicating that Titan's ice I shell may be currently conductive \citep{Nimmo2010}. Possibly, part of this shell is clathrate hydrate, which has different transport properties. If the mantle of high-pressure ice is convecting, then a system of equations, for the convective instability of and temperature contrasts across the thermal boundary layers at the interface with the ice/rock interior, is necessary to precisely calculate the thermal evolution of Titan. Because our focus is on the thermal evolution of Titan's deep interior, we assume for simplicity that $T_b$ is constant in all simulations.

\section{Numerical models}

We performed simulations to calculate the thermal history of Titan's ice/rock interior for 4.5~Gyr. The model described above was numerically iterated with a time step of 1~Myr. Internal temperatures were compared to the phase diagram for pure water ice from \citet{Petrenko1999} at each time step to determine whether melting occurs. Titan, as a whole, has $\bar\rho$ = 1,881~kg~m$^{-3}$ and $S$ $\approx$ 0.44. We assume that Titan is initially composed of an outer shell with a thickness of 575~km and a density of 1,200~kg~m$^{-3}$. Therefore, $R_i$ = 2,000~km and $S$ $\approx$ 0.82 in the ice/rock interior. The average density of the ice/rock interior, $\bar\rho$ = 2,653~kg~m$^{-3}$. In this simple two-layer model, the normalized moment of inertia coefficient $C$ $\approx$ 0.343. The melting temperatures of water ice are $\sim$280 and 500~K at pressures of $\sim$0.7 and 4.7~GPa, respectively, which are the conditions at the top and bottom of the ice/rock interior.

A wide parameter space was explored by varying the initial conditions for hundreds of simulations. The temperature at the top of the thermal boundary layer in the ice/rock interior was fixed at $T_b$ = 200~K, but we used both $T_m(0)$ = 200 and 125~K to explore the effects of different initial temperature gradients, as suggested by models of Titan's accretion \citep{Mitri2010}. Initial rock mass fraction gradients were imposed that produced effective density changes throughout the ice/rock interior as large as $\Delta\rho_s\sim0.1\bar\rho$ (i.e., $\sim$10$\%$). An initial temperature gradient, if present, also produces an effective density change of $\Delta\rho_t\sim\alpha\Delta T_i$, where $\Delta T_i$ = $T_b - T_m(0)$. In some simulations, no initial compositional gradient was imposed. Because of the large uncertainty on the viscosity of ice/rock mixtures, we tested effective viscosities near melting in the range $\nu_0$ = 10$^{11}$ to 10$^{15}$~Pa~s. (Even larger values increase the likelihood of melting and lower values are much lower than physically plausible.) To account for the possibility of potassium leaching and uncertainties in the rock mineralogy, we conducted simulations using three different depletion factors: $f$~=~1.0, 0.7, and 0.3. Guided by the results of these simulations, we construct some plausible present-day internal structure models for Titan.

\section{Results}

The following sections summarize the results of our simulations, beginning with the presentation of four representative thermal histories. Next, we discuss the effects of various initial conditions using a large number of simulation results. Finally, simulations in which melting occurred are analyzed to determine the details of melting in the ice/rock interior. 

\subsection{Sample thermal histories}

\begin{figure*}
\centering
\noindent\includegraphics[width=30pc]{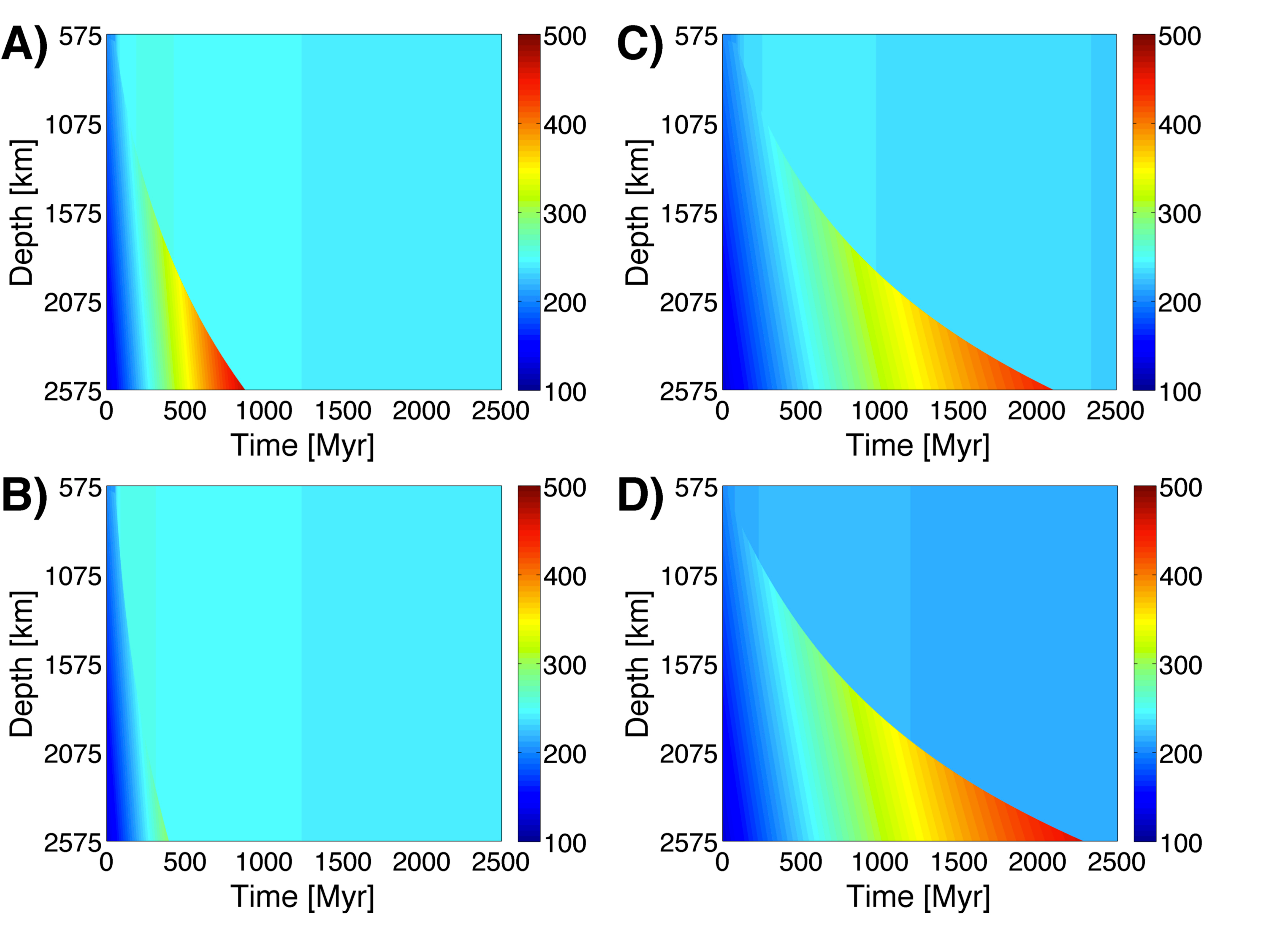}
\caption{Representative results for simulations in which melting did not occur. Temperature profiles in Titan's ice/rock interior are plotted until the time after which the convecting interior cools monotonically. The thermal boundary layers on either side of the conductive/convective interface are neither modeled nor shown. Default initial conditions for the four simulations are $T_b$~= 200~K, $T_m(0)$~= 125~K,  $f$~= 1, $\log_{10}(\nu_0)$~= 14, and $\Delta\rho_s$~= $0.01\bar\rho$. Panel B has $\Delta\rho_s$~= $0$. Panel C has $f$~= 0.3. Panel D has $f$~= 0.3 and $\log_{10}(\nu_0)$~= 12.} \label{fig:color}
\end{figure*}

Four examples of calculated thermal histories are shown in Fig.~\ref{fig:color}. In each simulation, a convecting layer with constant temperature grows over time, eventually encompassing the entire ice/rock interior. From 2.5 to 4.5~Gyr, this convecting ice/rock layer cools monotonically. Underneath the convecting layer, temperatures increase from radiogenic heating, although the initial linear temperature gradient is preserved. The internal temperature profile must be continuous; our model implies the existence of thermal boundary layers on both sides of the conductive/convective interface. These relatively thin layers are neither explicitly modeled nor depicted in Fig.~\ref{fig:color}. No melting occurs in any of these four examples within 4.5~Gyr because the convecting layer grows fast enough to encompass the entire interior before temperatures in the non-convecting layer increase enough to cause widespread melting and differentiation.

Comparisons between pairs of these simulations illuminate the effects of varying the magnitude of compositional gradients, potassium depletion, and initial temperature gradients. Simulations shown in panels A and B, for instance, both have $f$ = 1.0, $\nu_0$ = 10$^{14}$~Pa~s, and $T_m(0)$ = 125~K. Panel A features a $\sim$$1\%$ compositional gradient, i.e., $\Delta\rho_s = 0.01\bar\rho$, whereas the initial density gradient in panel B is only the result of the initial temperature gradient. The growth of the convecting layer is relatively delayed by $\sim$500~Myr in panel A, allowing an additional $\sim$100~K of heating in the ice/rock interior.

Panels A and C illustrate the effects of potassium depletion on the thermal evolution of an ice/rock interior with a $\sim$1$\%$ compositional gradient. All initial conditions are identical for both simulations, except $f$ = 0.3 in panel C. The primary effect of potassium depletion is to slow the growth of the convecting layer because heating in the underlying, non-convecting layer is necessary to overcome the initial compositional gradient. Specifically, convection in the entire ice/rock interior is delayed by $\sim$1.25~Gyr. The maximum temperatures reached in the non-convecting ice/rock interior, however, are only slightly lower in panel C than in panel A. On the other hand, the temperature of the convecting layer is generally reduced in panel C, since the convecting layer is relatively thin, internal heating is lessened, and the incorporation of hot material from the non-convecting interior is more gradual.

Finally, panels C and D reveal the effects of varying the effective viscosity near melting. Both simulations include substantial potassium depletion ($f$ = 0.3) and start with a $\sim$1$\%$ compositional gradient. While panel C features $\nu_0$ = 10$^{14}$~Pa~s as usual, $\nu_0$ = 10$^{12}$~Pa~s in panel D. With a lower viscosity, the convecting layer has a lower temperature and takes slightly longer to encompass the entire interior. In fact, convecting layer growth is faster for the first $\sim$500~Myr in panel D, but marginally slower thereafter. Thus, heat flux out of the convecting layer is relatively low after a period of relatively rapidly cooling and convecting layer growth. Compared to the other initial conditions, however, the sensitivity of the model to uncertainties in viscosity is very small.
 
\subsection{Sensitivity to initial conditions}

\begin{figure*}
\noindent\includegraphics[width=39pc]{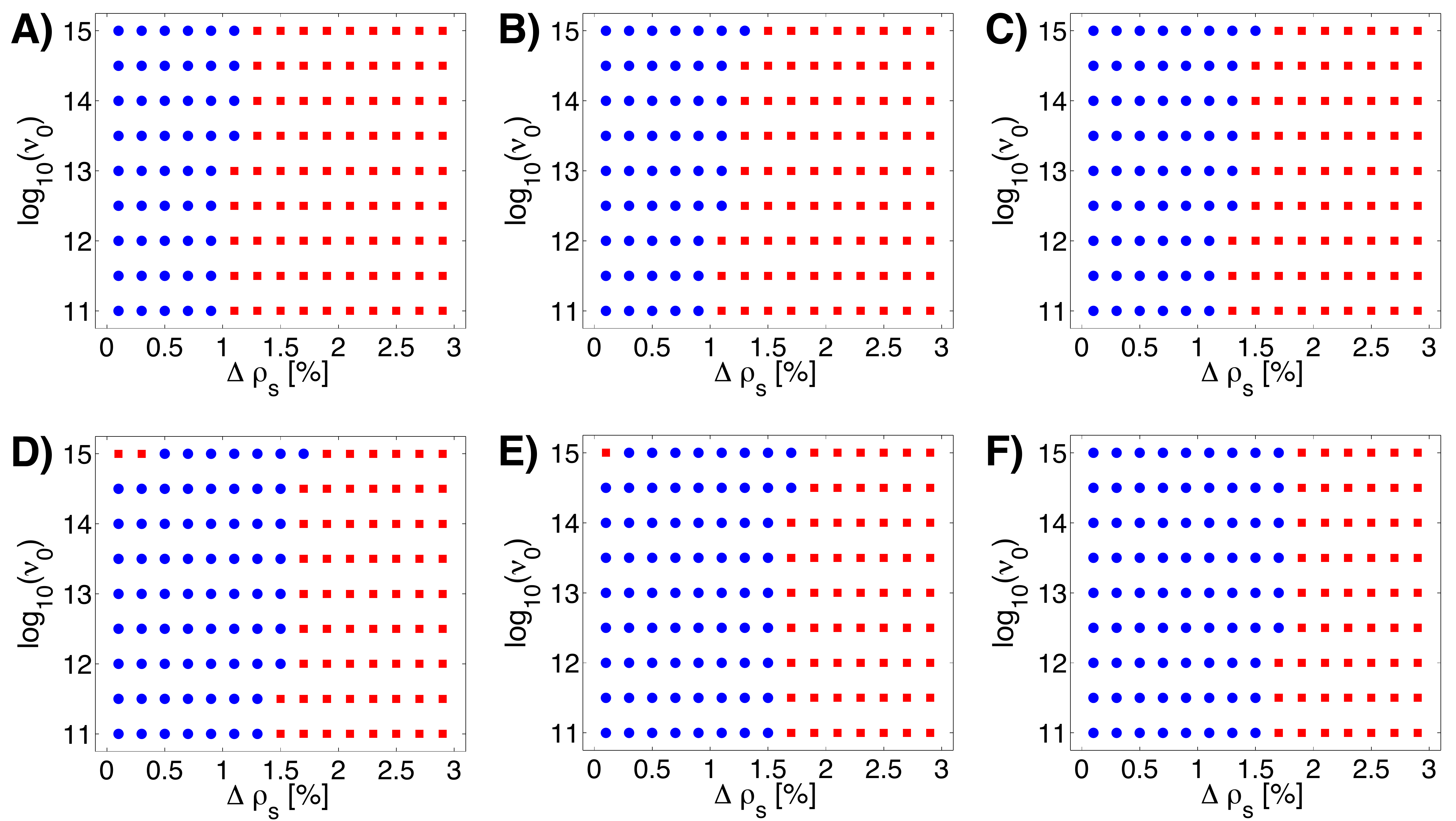}
\caption{Sensitivity analyses results for 810 simulations exploring the effects of potassium depletion, effective viscosity near melting, and initial temperature and compositional gradients. In each panel, the horizontal axis shows the total density variation across the ice/rock interior produced by the initial compositional gradient. The vertical axes show the dynamic viscosity near melting assumed for each simulation. Blue circles represent simulations in which no melting occurred in 4.5 Gyr; red squares signify simulations in which melting occurred at any point in the ice/rock interior. All simulations have $T_b$ = 200 K. From left to right, each column of panels features $f$ = 1.0, 0.7, and 0.3. The top and bottom rows of panels have $T_m(0)$ = 125 and 200 K, respectively. For initial compositional gradients producing $\Delta\rho_s$ $>$ $0.02\bar\rho$, melting occurs during all simulations.} \label{fig:melt}
\end{figure*}

Figure~\ref{fig:melt} shows results from 810 simulations of the thermal evolution of Titan's ice/rock interior. The purpose of each panel is to illustrate whether or not melting occurs for a particular set of initial conditions. Horizontal axes indicate the range of density differences across the ice/rock interior induced by an initial compositional gradient, i.e., $\Delta\rho_s$~= 0 to 0.03$\bar\rho$. Each vertical axis represents a range of effective viscosities near melting, i.e., $\nu_0$~= 10$^{11}$ to 10$^{15}$~Pa~s. Blue circles signify simulations in which melting never occurred, whereas red squares represent simulations where widespread melting would have caused differentiation of Titan's ice/rock interior within 4.5~Gyr.

Each panel represents a series of simulations that started with different initial temperature gradients and magnitudes of potassium depletion. The top and bottom rows of panels, for instance, have $T_m(0)$~= 125 and 200~K, respectively. In general, melting is less likely to occur without an initial temperature gradient, because a temperature decrease with depth induces a density gradient that suppresses convection. High-viscosity simulations with low compositional gradients in panels D and E feature melting, while the corresponding simulations in panels A and B do not. The lack of a significant density gradient from compositional or temperature changes in these simulations causes the convecting layer to very quickly encompass the interior, homogenizing internal temperatures. Because the average internal temperature is comparatively high at first, however, melting occurs near the top of the ice/rock interior, where melting temperatures are relatively low.

Figure~\ref{fig:melt} also elucidates the effects of potassium depletion. From left to right, columns of panels have $f$ = 1.0 (A and D), 0.7 (B and E), and 0.3 (C and F). Reducing the abundance of potassium in the rock component of the ice/rock interior only makes melting marginally less likely. As seen in the example thermal histories, potassium depletion only lengthens the time scales on which temperatures increase; the maximum temperatures reached in ice/rock interior are barely lessened. For any plausible choice of initial conditions, melting occurs if the initial compositional gradient is larger than $\sim$2$\%$.

\subsection{When and where melting occurs}

\begin{figure}
\noindent\includegraphics[width=20pc]{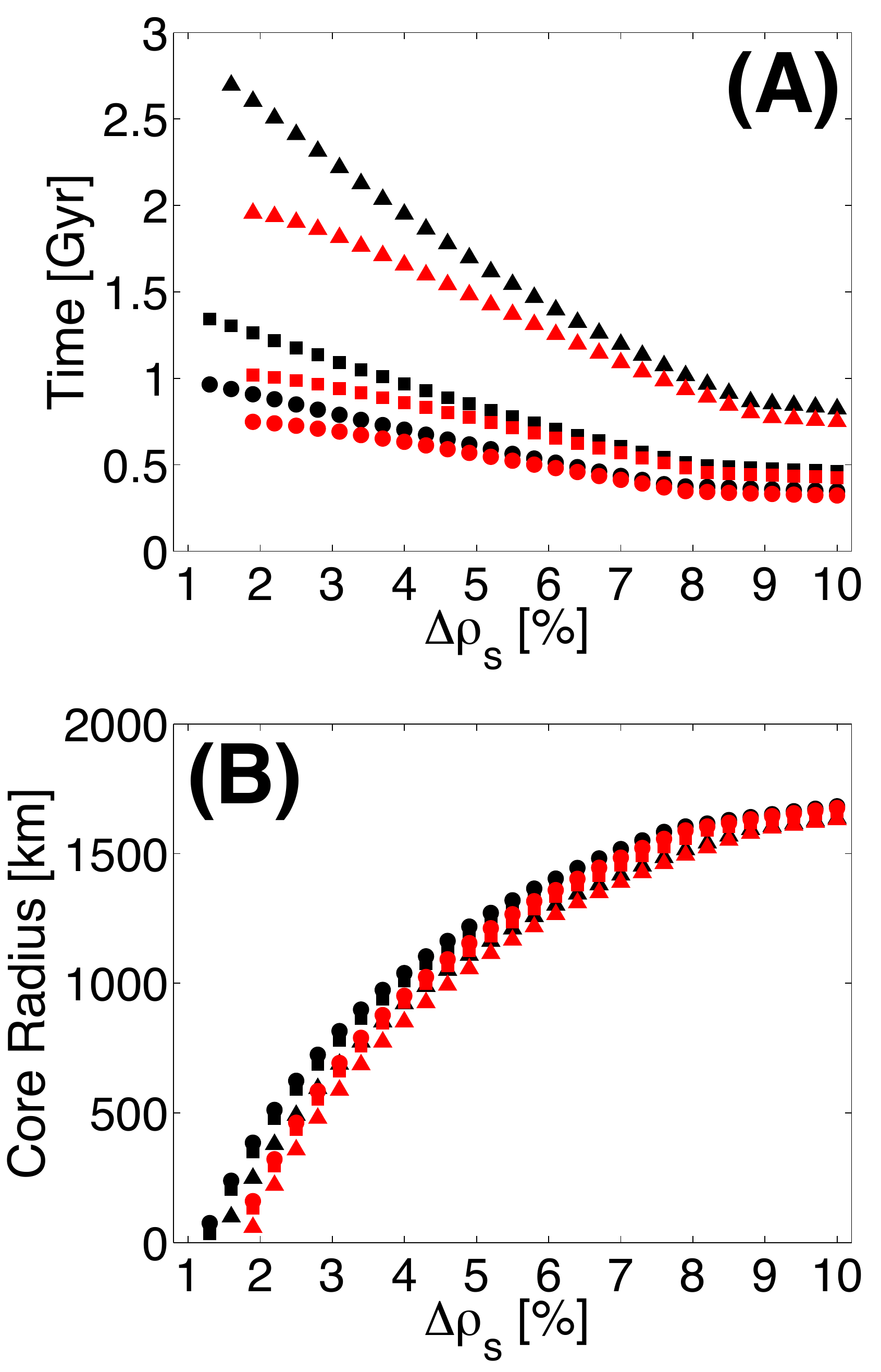}
\caption{Details of melting during 186 simulations of Titan's thermal evolution with initial compositional gradients that produce density contrasts from 0.01$\bar\rho$ to 0.1$\bar\rho$ across the ice/rock interior. Panel A shows when melting occurs in the non-convecting ice/rock interior. Panel B illustrates the size of the non-convecting ice/rock core at the time when melting occurs. All simulations have $T_b$ = 200 K. Black and red symbols respectively represent $T_m(0)$ = 125 and 200 K. Circular, square, and triangular markers represent $f$ = 1.0, 0.7, and 0.3, respectively.} \label{fig:t&d}
\end{figure}

Figure~\ref{fig:t&d} contains more information about 173 simulations in which melting occurred for initial compositional gradients ranging from 1 to 10$\%$. Specifically, panel A illustrates the time at which melting occurred and panel B shows the value of $R_c$, the radius of the non-convecting ice/rock interior, at the time when melting began. As in Fig.~\ref{fig:melt}, six different sets of initial conditions were used. That is, black and red symbols respectively represent $T_m(0)$ = 125 and 200~K. For circular, square, and triangular markers, respectively, $f$ = 1.0, 0.7, and 0.3. With large density gradients, the non-convecting ice/rock core always melts before the ice in the convecting layer. 

The magnitudes of potassium depletion and the initial temperature gradient more strongly affect when melting occurs than where it begins. That is, increasing potassium depletion can delay melting by $\sim$500~Myr to 1.75~Gyr for a given initial compositional gradient, but the radius of the non-convecting core at the time when melting begins only slightly decreases. With no initial temperature gradient but a higher average internal temperature, melting occurs sooner, but the convecting layer has grown further before melting begins. As compositional gradients increase, melting occurs sooner, the convecting layer is thinner when melting occurs, and the simulation's sensitivity to the choice of initial conditions decreases. Beginning the simulations at $t_0$ = 5 or 10~Myr after the origin of the CAIs delayed the onset of melting by $\ll$100~Myr.

\section{Discussion}

\subsection{Stability of a partially differentiated Titan}

A partially differentiated Titan is stable over geologic time if it can safeguard the ice in its deep interior from widespread melting caused by radiogenic heating. If thermal convection is vigorous throughout the entire ice/rock layer, then melting can be avoided. Even for small compositional gradients ($\sim$1-1.5$\%$) a convecting layer can grow to encompass the entire ice/rock interior before internal temperature profiles cross the melting curve of water ice. Larger compositional gradients ($\sim$5-10$\%$), however, are plausible since Titan likely accreted from heterogenous planetesimals. Thermal buoyancy simply cannot overcome these stabilizing gradients without causing ice melting and complete differentiation. The likely presence of ammonia in the ice component, which may significantly lower the melting temperature \citep[e.g.,][]{Loveday2009}, compounds the problem.

No plausible selection of initial conditions allows sufficiently efficient heat removal in the presence of a large compositional gradient. A substantial amount of heat-producing $^{40}$K may be leached out of the rock component through contact with liquid water or ammonia-water solutions. Roughly 30$\%$ of the original abundance of potassium has been estimated to have been extracted from the silicates of Ganymede \citep{Kirk1987} and Titan \citep{Engel1994}, and almost complete leaching of several elements has been suggested for Enceladus \citep{Glein2010}. A reduction in the magnitude of radiogenic heating may be important to preventing Titan's interior from reaching temperatures associated with silicate dehydration \citep{Castillo2010}, but, as seen in Fig.~\ref{fig:Q}, any plausible abundance of radioisotopes, even with $\sim$70$\%$ leaching of $^{40}$K, will eventually produce enough heat to melt ice. Crucially, internal heating drives the growth of the double-diffusive convecting layer, so potassium depletion suppresses the initialization of convection throughout the entire ice/rock interior.

Double-diffusive convection involves opposing gradients of two components that have different diffusivities. With components like salt and water, both diffusivities are non-zero, and multiple convecting layers form. In Titan's ice/rock system, however, the diffusivity of the rock component is effectively zero, and we only model the growth of a single layer. For small compositional gradients, the time scale for the growth of a convective instability of a new layer is much greater the time scale on which the first layer grows. With sufficiently large compositional gradients, growth of the convecting layer is slow enough that additional layers could form if melting never occurred, but melting typically occurs in $\sim$250 to 750~Myr. In any case, heat transport through a series of convecting layers is comparatively inefficient \citep[e.g.,][]{Turner1974}, and the formation of many layers is unlikely to prevent complete differentiation. 

\subsection{Alternative internal structures}

A partially differentiated Titan is a non-unique interpretation of available gravity data \citep{Iess2010}. Because Titan's ice/rock interior is unstable to differentiation over geologic time, alternate models should be considered. One popular hypothesis is that Titan's interior is chiefly composed of hydrated silicates \citep[e.g.,][]{Fortes2007,Castillo2010,Fortes2012}. Three serpentine polymorphs are typically found in carbonaceous chondrites: chrysotile, lizardite, and antigorite. Antigorite is the most abundant, and its properties are usually used to track the thermochemical evolution of model assemblages \citep[e.g.,][]{Castillo2010}. After heating melts the ice component of an ice/rock mixture, redistributing rock and melt into a stable density structure with homogenized internal temperatures takes a few hundred Myr \citep{Kirk1987,Lunine1987}. The serpentinization reaction between the liberated silicates and the surrounding liquid would likely proceed to completion well before core overturn is finished \citep{MacDonald1985}.

Models assuming a core composed entirely of antigorite are missing a significant mass of iron. The formula for the magnesium end-member of antigorite is Mg$_3$Si$_2$O$_5$(OH)$_4$. But Fe$^{2+}$ and Fe$^{3+}$ often substitute into the Mg$^{2+}$ and Al$^{3+}$ sites, respectively, yielding a typical iron number of Fe/(Fe+Mg) $\sim$ 0.2 \citep{Scott2002}. That is, antigorite has $<$14 wt$\%$ Fe, significantly less than the $\sim$18.2 wt$\%$ Fe typically found in CI chondrites \citep[p.~29]{Hutchinson2004}. So, $\sim$4 wt$\%$ Fe, equivalent to an Fe-rich core with a radius of a few hundred km, must accompany a mass of antigorite created from the serpentinization of silicates with chondritic composition, although some iron may be in soluble form in Titan's ocean. The presence of a relatively small Fe-rich core does not dramatically affect Titan's moment of inertia \citep{Castillo2010}.

\begin{figure*}
\noindent\includegraphics[width=39pc]{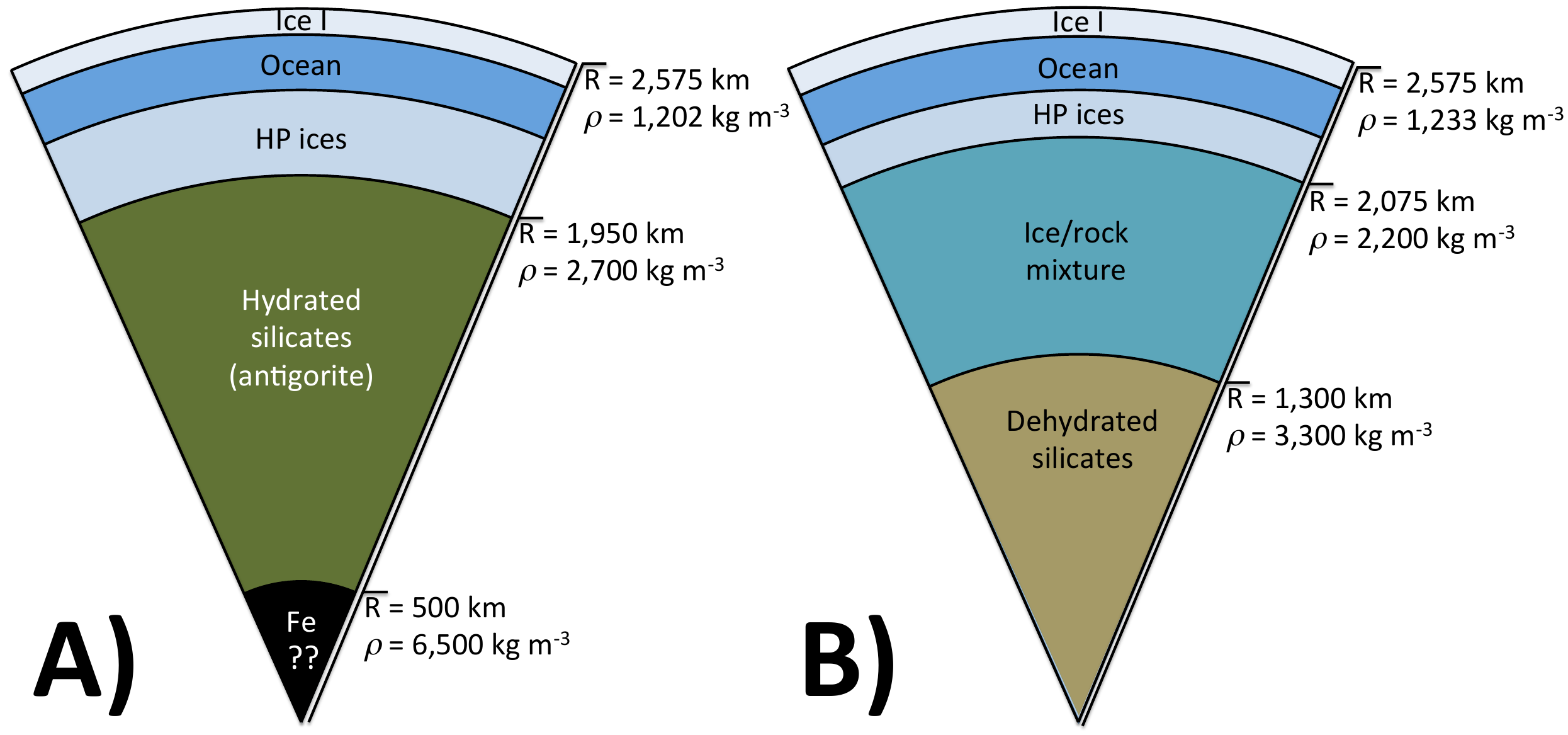}
\caption{Two end-member internal structures for Titan that have $C$ $\sim$ 0.34. Possible material layers include an Fe-rich core, dehydrated silicates, hydrated silicates, and an ice/rock mixture. Each model includes high-pressure ices, a water ocean with dissolved gases and solids, and a layer of water ice I, all of which are modeled as a single shell with constant density. Titan's atmosphere is neither depicted nor modeled.} \label{fig:Titans} \label{lastfigure}
\end{figure*}

Figure~\ref{fig:Titans} shows two end-member models for the present-day internal structure of Titan for which $C$ $\sim$ 0.34. In structure A, Titan is dominated by an hydrated silicate core with $R$~= 1,950~km and $\bar\rho$~= 2,700~kg~m$^{-3}$. An Fe-rich core with $R$~= 500~km and $\bar\rho$~= 6,500~kg~m$^{-3}$, representing $\sim$4 wt$\%$ of the iron/rock interior is also present, although the formation of an Fe-rich core would likely dehydrate the overlying silicates. The overlying shell, composed of an ice I lid, an ocean, and high-pressure ice polymorphs, is modeled as a single layer with an average density of 1,202 kg m$^{-3}$. The lower $\sim$300~km of antigorite could be replaced with dehydrated silicates in structure A without decreasing the MoI coefficient below $C\sim0.332$

A key question is whether significant dehydration of the silicate core should be expected. The equilibrium temperature for the antigorite dehydration reaction is $\sim$800 to 900~K for conditions appropriate to Titan \citep[e.g.,][]{Perrillat2005}, although dehydration in a water saturated environment may be kinetically hindered and not occur until temperatures reach $\sim$1060~K \citep{Seipold2003}, and the heat flux from the advection of warm water could be significant. \citet{Castillo2010} found that conductive heat transport could delay significant dehydration until very recently, assuming $\sim$30$\%$ potassium depletion and a relatively high specific heat for antigorite. Serpentine dehydration at the present may be responsible for melting clathrate hydrates in the overlying ice shell, releasing methane \citep{Tobie2006} and radiogenic $^{40}$Ar \citep{Niemann2005} onto the surface and into the atmosphere. We suggest that, if Titan formed partially differentiated, double-diffusive convection in the initial ice/rock mixture may have delayed core overturn for as long as $\sim$1 Gyr beyond what was assumed in \citet{Castillo2010}, helping the eventual core of hydrated silicates avoid dehydration.

Current data do not exclude partially differentiated models featuring a large silicate core. Structure B in Fig.~\ref{fig:Titans} is an example of such a model, similar to those considered in \citet{McKinnon2011}. An ice/rock mantle with $R$ = 2,075~km, $\bar\rho$ = 2,200~kg~m$^{-3}$, and $S$ $\sim$ 0.63 overlies a core of dehydrated silicates with $R$ = 1,300~km and $\bar\rho$ = 3,300~kg~m$^{-3}$. A decreased rock mass fraction and a smaller length scale for the ice/rock layer might allow efficient convection, even in the presence of a stabilizing compositional gradient. Evaluating the thermal stability of such a structure would require coupling the thermal evolution of all three layers, potentially including the effects of tidal dissipation, as has been done for Ganymede \citep{Bland2009}. More complicated variations on structure B featuring different ice/rock mixtures or a hydrated portion of the silicate core are also valid interpretations of the gravity data. Differences in the predicted tidal responses of these two models may allow discrimination between these two alternatives \citep{Castillo2012}.

\subsection{Implications for other icy satellites}

Gravity measurements by spacecraft suggest a spectrum of internal structures for the icy satellites of Jupiter and Saturn. The formation of an undifferentiated or a partially differentiated satellite requires a sequence of conditions: slow accretion in a gas-starved disk, delayed to avoid catastrophic heating from the radioactive decay of short-lived isotopes, and then survival of intense impacts from the Late Heavy Bombardment \citep{Barr2008,BarrCanup2010,Barr2010}. The Radau-Darwin approximation is primarily used to calculate the MoI of a satellite with the measured $J_2$ and $C_{22}$ coefficients, but a unique solution only exists for a satellite in hydrostatic equilibrium.

Non-hydrostatic structures in the degree 2 gravity of Titan may introduce significant inaccuracies into the Radau-Darwin approximation \citep{Gao2012}. Indeed, \citet{Iess2010} reported non-hydrostatic geoid height variations as large as 19~m for Titan, possibly indicating that Titan's MoI coefficient could closer to that of Ganymede than earlier reported. The measured MoI for Callisto is likewise subject to large uncertainties because of possible non-hydrostatic effects, which are not excluded by current data \citep{Anderson2001}, but an equivalent departure from hydrostatic equilibrium would have a relatively small effect on Ganymede. The Galilean satellites are not as compositionally heterogeneous as the major Saturnian satellites \citep{Anderson1996Io,Anderson1996,Anderson1998,Anderson2001}, so the likelihood of substantial compositional gradients suppressing convection in Callisto may be lessened. In any case, the radio science instrument on the planned \textit{JUICE} (JUpiter ICy moons Explorer) mission may allow the vastly improved measurement of the gravity fields of Ganymede and Callisto and thus the identification of gravity anomalies and significant non-hydrostatic features. With these new data, we may continue to vet the spectrum of internal differentiation states proposed for icy satellites.

The internal structures of Saturn's medium-sized satellites may also shed light on the formation and evolution of the Saturnian system. Unfortunately, their MoIs are not yet tightly constrained. Gravity data from the Cassini mission suggest that Rhea, Saturn's second-largest satellite, has a larger MoI coefficient than Titan \citep{Iess2007}. Based on a preliminary analysis, \citet{Anderson2007} claimed that Rhea's interior was a homogenous ice/rock mixture (i.e., with no compositional gradient). Further analysis indicates that rock must be at least somewhat concentrated towards the center of Rhea \citep{Iess2007}, perhaps in a core with relatively high rock mass fraction, even if Rhea's gravity field is subject to significant non-hydrostatic effects \citep{Mackenzie2008}. The rock mass fraction of Rhea is almost half that of Titan, so a relative scarcity of radiogenic heating may stave off differentiation. The simple model developed in this paper for Titan may not be directly applicable to Rhea, which is much smaller than Titan, but future thermal evolution simulations would help evaluate whether the proposed internal structures are plausible.

\section{Conclusions}

Titan can survive accretion and the Late Heavy Bombardment without suffering complete differentiation if it formed slowly in a gas-starved disk, and a partially differentiated Titan is consistent with recently obtained gravity data. But radiogenic heating tends to melt ice in the undifferentiated interior over geologic time, causing irreversible sinking of the rock component. If the ice/rock interior is initially homogenous, then thermal convection may remove sufficient heat to prevent melting. Two sources of compositional gradients, however, act to suppress convection. First, Titan likely had a temperature profile immediately after accretion in which temperature decreased with depth. Second, accretion from heterogenous planetesimals would produce a stabilizing rock mass fraction gradient, where rock mass fraction and thus density increased with depth. Convection in the presence of opposing compositional and thermal gradients is ``double-diffusive", comprising a convecting layer that slowly grows from the top of the ice/rock interior. For compositional gradients larger than a few percent, this convecting layer does not grow fast enough to prevent widespread melting in the non-convecting core. A partially differentiated Titan is therefore unstable over geologic time. The general instability of ice/rock mixtures with compositional gradients may have significant consequences for other icy satellites.


\section{Acknowledgments}
J.G. O'Rourke thanks A.C. Barr, W.B. McKinnon, and I. Mosqueira for helpful discussions and the NASA Planetary Geology $\&$ Geophysics Undergraduate Research Program for support. Constructive comments from one reviewer improved the content and clarity of this manuscript. J.G. OÕRourke is supported by a National Science Foundation Graduate Research Fellowship.

\label{lastpage}

\section{References Cited}
\bibliographystyle{model2-names}

\begin{thebibliography}{77}
\expandafter\ifx\csname natexlab\endcsname\relax\def\natexlab#1{#1}\fi
\expandafter\ifx\csname url\endcsname\relax
  \def\url#1{\texttt{#1}}\fi
\expandafter\ifx\csname urlprefix\endcsname\relax\def\urlprefix{URL }\fi
\providecommand{\eprint}[2][]{\url{#2}}
\providecommand{\bibinfo}[2]{#2}
\ifx\xfnm\relax \def\xfnm[#1]{\unskip,\space#1}\fi
%Type = Article
\bibitem[{Anderson and Schubert(2007)Anderson and Schubert}]{Anderson2007}
\bibinfo{author}{Anderson, J.D.},
  \bibinfo{author}{Schubert, G.}, \bibinfo{year}{2007}.
\newblock \bibinfo{title}{{Saturn's satellite Rhea is a homogeneous mix of rock and ice}}.
\newblock \bibinfo{journal}{Geophys. Res. Lett.} \bibinfo{volume}{37},
  \bibinfo{pages}{L02202, doi:10.1029/2006GL028100}.
%Type = Article
\bibitem[{Anderson et~al.(2001)Anderson, Jacobson, McElrath, Moore, Schubert
  and Thomas}]{Anderson2001}
\bibinfo{author}{Anderson, J.D.}, \bibinfo{author}{Jacobson, R.A.},
  \bibinfo{author}{McElrath, T.P.}, \bibinfo{author}{Moore, W.B.},
  \bibinfo{author}{Schubert, G.}, \bibinfo{author}{Thomas, P.C.},
  \bibinfo{year}{2001}.
\newblock \bibinfo{title}{Shape, mean radius, gravity field, and interior
  structure of {Callisto}}.
\newblock \bibinfo{journal}{Icarus} \bibinfo{volume}{153},
  \bibinfo{pages}{157--161, doi:10.1006/icar.2001.6664}.
%Type = Article
\bibitem[{Anderson et~al.(1996a)Anderson, Lau, Sjogren, Schubert and
  Moore}]{Anderson1996}
\bibinfo{author}{Anderson, J.D.}, \bibinfo{author}{Lau, E.L.},
  \bibinfo{author}{Sjogren, W.L.}, \bibinfo{author}{Schubert, G.},
  \bibinfo{author}{Moore, W.B.}, \bibinfo{year}{1996}a.
\newblock \bibinfo{title}{Gravitational constraints on the internal structure
  of {Ganymede}}.
\newblock \bibinfo{journal}{Nature} \bibinfo{volume}{384},
  \bibinfo{pages}{541--543}.
%Type = Article
\bibitem[{Anderson et~al.(1998)Anderson, Schubert, Jacobson, Lau, Moore and
  Sjogren}]{Anderson1998}
\bibinfo{author}{Anderson, J.D.}, \bibinfo{author}{Schubert, G.},
  \bibinfo{author}{Jacobson, R.A.}, \bibinfo{author}{Lau, E.L.},
  \bibinfo{author}{Moore, W.B.}, \bibinfo{author}{Sjogren, W.L.},
  \bibinfo{year}{1998}.
\newblock \bibinfo{title}{Europa's differentiated internal structure:
  {Inferences from four Galileo encounters}}.
\newblock \bibinfo{journal}{Science} \bibinfo{volume}{281},
  \bibinfo{pages}{2019--2022}.
%Type = Article
\bibitem[{Anderson et~al.(1996b)Anderson, Sjogren and
  Schubert}]{Anderson1996Io}
\bibinfo{author}{Anderson, J.D.}, \bibinfo{author}{Sjogren, W.L.},
  \bibinfo{author}{Schubert, G.}, \bibinfo{year}{1996}b.
\newblock \bibinfo{title}{{Galileo gravity results and the internal structure
  of Io}}.
\newblock \bibinfo{journal}{Science} \bibinfo{volume}{272},
  \bibinfo{pages}{709--712}.
  %Type = Article
\bibitem[{Asphaug and Reufer(2013)Asphaug and Reufer}]{Asphaug2013}
\bibinfo{author}{Asphaug, E.}, \bibinfo{author}{Reufer, A.},
  \bibinfo{year}{2013}.
\newblock \bibinfo{title}{{Late origin of the {Saturn} system}}.
\newblock \bibinfo{journal}{Icarus} \bibinfo{volume}{223},
  \bibinfo{pages}{544--565, doi:10.1016/j.icarus.2012.12.009}.
%Type = Article
\bibitem[{Atreya et~al.(2006)Atreya, Adams, Niemann, Demick-Montelara, Owen,
  Fulchignoni, Ferri and Wilson}]{Atreya2006}
\bibinfo{author}{Atreya, S.K.}, \bibinfo{author}{Adams, E.Y.},
  \bibinfo{author}{Niemann, H.B.}, \bibinfo{author}{Demick-Montelara, J.E.},
  \bibinfo{author}{Owen, T.C.}, \bibinfo{author}{Fulchignoni, M.},
  \bibinfo{author}{Ferri, F.}, \bibinfo{author}{Wilson, E.H.},
  \bibinfo{year}{2006}.
\newblock \bibinfo{title}{Titan's methane cycle}.
\newblock \bibinfo{journal}{Planet. Space Sci.} \bibinfo{volume}{54},
  \bibinfo{pages}{1177--1187, doi:10.1016/j.pss.2006.05.028}.
%Type = Article
\bibitem[{Barr and Canup(2008)}]{Barr2008}
\bibinfo{author}{Barr, A.C.}, \bibinfo{author}{Canup, R.M.},
  \bibinfo{year}{2008}.
\newblock \bibinfo{title}{Constraints on gas giant satellite formation from the
  interior states of partially differentiated satellites}.
\newblock \bibinfo{journal}{Icarus} \bibinfo{volume}{198},
  \bibinfo{pages}{163--177, doi:10.1016/j.icarus.2008.07.004}.
%Type = Article
\bibitem[{Barr and Canup(2010)}]{BarrCanup2010}
\bibinfo{author}{Barr, A.C.}, \bibinfo{author}{Canup, R.M.},
  \bibinfo{year}{2010}.
\newblock \bibinfo{title}{Origin of the {Ganymede-Callisto} dichotomy by
  impacts during the late heavy bombardment}.
\newblock \bibinfo{journal}{Nat. Geosci.} \bibinfo{volume}{3},
  \bibinfo{pages}{164--167, doi:10.1038/NGEO746}.
%Type = Article
\bibitem[{Barr et~al.(2010)Barr, Citron and Canup}]{Barr2010}
\bibinfo{author}{Barr, A.C.}, \bibinfo{author}{Citron, R.I.},
  \bibinfo{author}{Canup, R.M.}, \bibinfo{year}{2010}.
\newblock \bibinfo{title}{Origin of a partially differentiated {Titan}}.
\newblock \bibinfo{journal}{Icarus} \bibinfo{volume}{209},
  \bibinfo{pages}{858--862, doi:10.1016/j.icarus.2010.05.028}.
%Type = Article
\bibitem[{Barr and McKinnon(2007)}]{Barr2007}
\bibinfo{author}{Barr, A.C.}, \bibinfo{author}{McKinnon, W.B.},
  \bibinfo{year}{2007}.
\newblock \bibinfo{title}{{Convection in ice I shells and mantles with
  self-consistent grain size}}.
\newblock \bibinfo{journal}{J. Geophys. Res.} \bibinfo{volume}{112},
  \bibinfo{pages}{E02012, doi:10.1029/2006JE002781}.
%Type = Article
\bibitem[{Bills and Nimmo(2011)}]{Bills2011}
\bibinfo{author}{Bills, B.G.}, \bibinfo{author}{Nimmo, F.},
  \bibinfo{year}{2011}.
\newblock \bibinfo{title}{{Rotational dynamics and internal structure of
  Titan}}.
\newblock \bibinfo{journal}{Icarus} \bibinfo{volume}{214},
  \bibinfo{pages}{351--355, doi:10.1016/j.icarus.2011.04.028}.
%Type = Article
\bibitem[{Bland et~al.(2009)Bland, Showman and Tobie}]{Bland2009}
\bibinfo{author}{Bland, M.T.}, \bibinfo{author}{Showman, A.P.},
  \bibinfo{author}{Tobie, G.}, \bibinfo{year}{2009}.
\newblock \bibinfo{title}{The orbital-thermal evolution and global expansion of
  {Ganymede}}.
\newblock \bibinfo{journal}{Icarus} \bibinfo{volume}{200},
  \bibinfo{pages}{207--221, doi:10.1016/j.icarus.2008.11.016}.
%Type = Article
\bibitem[{Canup(2010)}]{Canup2010}
\bibinfo{author}{Canup, R.M.}, \bibinfo{year}{2010}.
\newblock \bibinfo{title}{Origin of {Saturn's} rings and inner moons by mass
  removal from a lost {Titan-sized} satellite}.
\newblock \bibinfo{journal}{Nature} \bibinfo{volume}{468},
  \bibinfo{pages}{943--946}.
%Type = Article
\bibitem[{Canup and Ward(2002)}]{Canup2002}
\bibinfo{author}{Canup, R.M.}, \bibinfo{author}{Ward, W.R.},
  \bibinfo{year}{2002}.
\newblock \bibinfo{title}{Formation of the {Galilean} satellites: conditions of
  accretion}.
\newblock \bibinfo{journal}{Astronom. J.} \bibinfo{volume}{124},
  \bibinfo{pages}{3404--3423}.
%Type = Article
\bibitem[{Castillo-Rogez and Lunine(2010)}]{Castillo2010}
\bibinfo{author}{Castillo-Rogez, J.C.}, \bibinfo{author}{Lunine, J.I.},
  \bibinfo{year}{2010}.
\newblock \bibinfo{title}{Evolution of {Titan's} rocky core constrained by
  {Cassini} observations}.
\newblock \bibinfo{journal}{Geophys. Res. Lett.} \bibinfo{volume}{37},
  \bibinfo{pages}{L20205, doi:10.1029/2010GL044398}.
%Type = Article
\bibitem[{Castillo-Rogez and Lunine(2012)}]{Castillo2012}
\bibinfo{author}{Castillo-Rogez, J.C.}, \bibinfo{author}{Lunine, J.I.},
  \bibinfo{year}{2012}.
\newblock \bibinfo{title}{{Tidal response of Titan's interior models consistent
  with Cassini-derived constraints}}.
\newblock \bibinfo{journal}{LPSC XLIII Abstracts $\#$1707} .
%Type = Article
\bibitem[{Engel et~al.(1994)Engel, Lunine and Norton}]{Engel1994}
\bibinfo{author}{Engel, S.}, \bibinfo{author}{Lunine, J.I.},
  \bibinfo{author}{Norton, D.L.}, \bibinfo{year}{1994}.
\newblock \bibinfo{title}{Silicate interactions with ammonia-water fluids on
  early {Titan}}.
\newblock \bibinfo{journal}{J. Geophys. Res.} \bibinfo{volume}{99},
  \bibinfo{pages}{3745--3752}.
%Type = Article
\bibitem[{Fortes(2012)}]{Fortes2012}
\bibinfo{author}{Fortes, A.D.}, \bibinfo{year}{2012}.
\newblock \bibinfo{title}{Titan's internal structure and the evolutionary
  consequences}.
\newblock \bibinfo{journal}{Planet. Space Sci.} \bibinfo{volume}{60},
  \bibinfo{pages}{10--17, doi:10.1016/j.pss.2011.04.010}.
%Type = Article
\bibitem[{Fortes et~al.(2007)Fortes, Grindrod, Trickett and
  Vocadlo}]{Fortes2007}
\bibinfo{author}{Fortes, A.D.}, \bibinfo{author}{Grindrod, P.M.},
  \bibinfo{author}{Trickett, S.K.}, \bibinfo{author}{Vocadlo, L.},
  \bibinfo{year}{2007}.
\newblock \bibinfo{title}{{Ammonium sulfate on Titan: Posible origin and role
  in cryovolcanism}}.
\newblock \bibinfo{journal}{Icarus} \bibinfo{volume}{188},
  \bibinfo{pages}{139--153, doi:10.1016/j.icarus.2006.11.002}.
%Type = Article
\bibitem[{Friedson and Stevenson(1983)}]{Friedson1983}
\bibinfo{author}{Friedson, A.J.}, \bibinfo{author}{Stevenson, D.J.},
  \bibinfo{year}{1983}.
\newblock \bibinfo{title}{Viscosity of rock-ice mixtures and applications to
  the evolution of icy satellites}.
\newblock \bibinfo{journal}{Icarus} \bibinfo{volume}{56},
  \bibinfo{pages}{1--14}.
%Type = Article
\bibitem[{Gao and Stevenson(2013)}]{Gao2012}
\bibinfo{author}{Gao, P.}, \bibinfo{author}{Stevenson, D.J.},
  \bibinfo{year}{2013}.
\newblock \bibinfo{title}{Nonhydrostatic effects and the determination of icy satellitesÕ moment of inertia}.
\newblock \bibinfo{journal}{Icarus}\bibinfo{volume}{226},
  \bibinfo{pages}{1185--1191, doi:10.1016/j.icarus.2013.07.034}.
  %Type = Article
\bibitem[{Gautier and Hersant(2005)}]{Gautier2005}
\bibinfo{author}{Gautier, D.}, \bibinfo{author}{Hersant, F.},
  \bibinfo{year}{2005}.
\newblock \bibinfo{title}{Formation and composition of planetesimals:
  {Trapping} volatiles by clathration}.
\newblock \bibinfo{journal}{Space Sci. Rev.} \bibinfo{volume}{116},
  \bibinfo{pages}{25--52, doi:10.1007/s11214--005--1946--2}.
%Type = Article
\bibitem[{Glein and Shock(2010)}]{Glein2010}
\bibinfo{author}{Glein, C.R.}, \bibinfo{author}{Shock, E.L.},
  \bibinfo{year}{2010}.
\newblock \bibinfo{title}{{Sodium chloride as a geophysical probe of a
  subsurface ocean on Enceladus}}.
\newblock \bibinfo{journal}{Geophys. Res. Lett.} \bibinfo{volume}{37},
  \bibinfo{pages}{L09204, doi:10.1029/2010GL042446,}.
%Type = Article
\bibitem[{Gomes et~al.(2005)Gomes, Levison, Tsiganis and
  Morbidelli}]{Gomes2005}
\bibinfo{author}{Gomes, R.}, \bibinfo{author}{Levison, H.F.},
  \bibinfo{author}{Tsiganis, K.}, \bibinfo{author}{Morbidelli, A.},
  \bibinfo{year}{2005}.
\newblock \bibinfo{title}{{Origin of the cataclysmic Late Heavy Bombardment
  period of the terrestrial planets}}.
\newblock \bibinfo{journal}{Nature} \bibinfo{volume}{435},
  \bibinfo{pages}{466--469, doi:10.1038/nature03676}.
%Type = Article
\bibitem[{Grindrod et~al.(2008)Grindrod, Fortes, Nimmo, Feltham, Brodholt and
  Vocadlo}]{Grindrod2008}
\bibinfo{author}{Grindrod, P.M.}, \bibinfo{author}{Fortes, A.D.},
  \bibinfo{author}{Nimmo, F.}, \bibinfo{author}{Feltham, D.L.},
  \bibinfo{author}{Brodholt, J.P.}, \bibinfo{author}{Vocadlo, L.},
  \bibinfo{year}{2008}.
\newblock \bibinfo{title}{{The long-term stability of a possible aqueous
  ammonium sulfate ocean inside Titan}}.
\newblock \bibinfo{journal}{Icarus} \bibinfo{volume}{197},
  \bibinfo{pages}{137--151, doi:10.1016/j.icarus.2008.04.006}.
%Type = Article
\bibitem[{Huppert and Turner(1981)}]{Huppert1981}
\bibinfo{author}{Huppert, H.E.}, \bibinfo{author}{Turner, J.S.},
  \bibinfo{year}{1981}.
\newblock \bibinfo{title}{Double-diffusive convection}.
\newblock \bibinfo{journal}{J. Fluid Mech.} \bibinfo{volume}{106},
  \bibinfo{pages}{299--329}.
%Type = Book
\bibitem[{Hutchinson(2004)}]{Hutchinson2004}
\bibinfo{author}{Hutchinson, R.}, \bibinfo{year}{2004}.
\newblock \bibinfo{title}{{Meteorites. A Petrologic, Chemical and Isotopic
  Synthesis}}.
\newblock \bibinfo{publisher}{Cambridge University Press}.
%Type = Article
\bibitem[{Iess et~al.(2012)Iess, Jacobson, Ducci, Stevenson, Lunine, Armstrong,
  Asmar, Racioppa, Rappaport and Tortora}]{Iess2012}
\bibinfo{author}{Iess, L.}, \bibinfo{author}{Jacobson, R.A.},
  \bibinfo{author}{Ducci, M.}, \bibinfo{author}{Stevenson, D.J.},
  \bibinfo{author}{Lunine, J.I.}, \bibinfo{author}{Armstrong, J.W.},
  \bibinfo{author}{Asmar, S.W.}, \bibinfo{author}{Racioppa, P.},
  \bibinfo{author}{Rappaport, N.J.}, \bibinfo{author}{Tortora, P.},
  \bibinfo{year}{2012}.
\newblock \bibinfo{title}{The tides of {Titan}}.
\newblock \bibinfo{journal}{Science} ,
  \bibinfo{pages}{doi:10.1126/science.1219631}.
%Type = Article
\bibitem[{Iess et~al.(2010)Iess, Rappaport, Jacobson, Racioppa, Stevenson,
  Tortora, Armstrong and Asmar}]{Iess2010}
\bibinfo{author}{Iess, L.}, \bibinfo{author}{Rappaport, N.J.},
  \bibinfo{author}{Jacobson, R.A.}, \bibinfo{author}{Racioppa, P.},
  \bibinfo{author}{Stevenson, D.J.}, \bibinfo{author}{Tortora, P.},
  \bibinfo{author}{Armstrong, J.W.}, \bibinfo{author}{Asmar, S.W.},
  \bibinfo{year}{2010}.
\newblock \bibinfo{title}{Gravity field, shape, and moment of intertia of
  {Titan}}.
\newblock \bibinfo{journal}{Science} \bibinfo{volume}{327},
  \bibinfo{pages}{1367--1369 , doi:10.1126/science.1182583}.
 %Type = Article
\bibitem[{Iess et~al.(2007)Iess, Rappaport, Tortora, Lunine, Armstrong, Asmar, Somenzi and Zingoni}]{Iess2007}
\bibinfo{author}{Iess, L.}, \bibinfo{author}{Rappaport, N.J.},
  \bibinfo{author}{Tortora, P.}, \bibinfo{author}{Lunine, J.},
  \bibinfo{author}{Armstrong, J.W.}, \bibinfo{author}{Asmar, S.W.},
  \bibinfo{author}{Somenzi, L.}, \bibinfo{author}{Zingoni, F.},
  \bibinfo{year}{2007}.
\newblock \bibinfo{title}{Gravity field and interior of {Rhea} from {Cassini} data analysis}.
\newblock \bibinfo{journal}{Icarus} \bibinfo{volume}{190},
  \bibinfo{pages}{585--593 , doi:10.1016/j.icarus.2007.03.027}.
%Type = Article
\bibitem[{Jacobson(2004)}]{Jacobson2004}
\bibinfo{author}{Jacobson, R.A.}, \bibinfo{year}{2004}.
\newblock \bibinfo{title}{The orbits of the major {Saturnian} satellites and
  the gravity field of {Saturn} from spacecraft and {Earth}-based
  observations}.
\newblock \bibinfo{journal}{Astrophys. J.} \bibinfo{volume}{128},
  \bibinfo{pages}{492--501, doi:10.1086/421738}.
%Type = Incollection
\bibitem[{Jaumann et~al.(2009)Jaumann, Kirk, Lorenz, Lopes, Stofan, Turtle,
  Keller, Wood, Sotin, Soderblom and Tomasko}]{Jaumann2009}
\bibinfo{author}{Jaumann, R.}, \bibinfo{author}{Kirk, R.L.},
  \bibinfo{author}{Lorenz, R.D.}, \bibinfo{author}{Lopes, R.M.C.},
  \bibinfo{author}{Stofan, E.}, \bibinfo{author}{Turtle, E.P.},
  \bibinfo{author}{Keller, H.U.}, \bibinfo{author}{Wood, C.},
  \bibinfo{author}{Sotin, C.}, \bibinfo{author}{Soderblom, L.A.},
  \bibinfo{author}{Tomasko, M.G.}, \bibinfo{year}{2009}.
\newblock \bibinfo{title}{Geology and surface processses on {Titan}}, in:
  \bibinfo{editor}{Brown, R.H.}, \bibinfo{editor}{Lebreton, J.P.},
  \bibinfo{editor}{Waite, J.H.} (Eds.), \bibinfo{booktitle}{{Titan from
  Cassini-Huygens}}. \bibinfo{publisher}{Springer}.
  chapter~\bibinfo{chapter}{5}.
%Type = Article
\bibitem[{Kirk and Stevenson(1987)}]{Kirk1987}
\bibinfo{author}{Kirk, R.L.}, \bibinfo{author}{Stevenson, D.J.},
  \bibinfo{year}{1987}.
\newblock \bibinfo{title}{Thermal evolution of a differentiated {Ganymede} and
  implications for surface features}.
\newblock \bibinfo{journal}{Icarus} \bibinfo{volume}{69},
  \bibinfo{pages}{91--134}.
%Type = Incollection
\bibitem[{Lissauer and Stevenson(2007)}]{Lissauer2007}
\bibinfo{author}{Lissauer, J.J.}, \bibinfo{author}{Stevenson, D.J.},
  \bibinfo{year}{2007}.
\newblock \bibinfo{title}{Formation of giant planets}, in:
  \bibinfo{editor}{Reipurth, B.}, \bibinfo{editor}{Jewitt, D.},
  \bibinfo{editor}{Keil, K.} (Eds.), \bibinfo{booktitle}{{Protostars and
  Planets V}}. \bibinfo{publisher}{University of Arizona Press},
  \bibinfo{address}{Tucson}, pp. \bibinfo{pages}{591--606}.
%Type = Article
\bibitem[{Loveday et~al.(2009)Loveday, Nelmes, Bull, Maynard-Casely and
  Guthrie}]{Loveday2009}
\bibinfo{author}{Loveday, J.S.}, \bibinfo{author}{Nelmes, R.J.},
  \bibinfo{author}{Bull, C.L.}, \bibinfo{author}{Maynard-Casely, H.E.},
  \bibinfo{author}{Guthrie, M.}, \bibinfo{year}{2009}.
\newblock \bibinfo{title}{{Observation of ammonia dihydrate in the AMH-VI
  structure at room temperature - possible implications for the outer solar
  system}}.
\newblock \bibinfo{journal}{High Pressure Res.} \bibinfo{volume}{29},
  \bibinfo{pages}{396--404, doi:10.1080/08957950903162057}.
%Type = Incollection
\bibitem[{Lunine et~al.(2009)Lunine, Choukroun, Stevenson and
  Tobie}]{Lunine2010}
\bibinfo{author}{Lunine, J.}, \bibinfo{author}{Choukroun, M.},
  \bibinfo{author}{Stevenson, D.J.}, \bibinfo{author}{Tobie, G.},
  \bibinfo{year}{2009}.
\newblock \bibinfo{title}{The origin and evolution of {Titan}}, in:
  \bibinfo{editor}{Brown, R.H.}, \bibinfo{editor}{Lebreton, J.P.},
  \bibinfo{editor}{Waite, J.H.} (Eds.), \bibinfo{booktitle}{{Titan from
  Cassini-Huygens}}. \bibinfo{publisher}{Springer}.
  chapter~\bibinfo{chapter}{3}.
%Type = Article
\bibitem[{Lunine and Stevenson(1987)}]{Lunine1987}
\bibinfo{author}{Lunine, J.I.}, \bibinfo{author}{Stevenson, D.J.},
  \bibinfo{year}{1987}.
\newblock \bibinfo{title}{Clathrate and ammonia hydrates at high pressure:
  {Application to the origin of methane on Titan}}.
\newblock \bibinfo{journal}{Icarus} \bibinfo{volume}{70},
  \bibinfo{pages}{61--77}.
%Type = Article
\bibitem[{MacDonald and Fyfe(1985)}]{MacDonald1985}
\bibinfo{author}{MacDonald, A.H.}, \bibinfo{author}{Fyfe, W.S.},
  \bibinfo{year}{1985}.
\newblock \bibinfo{title}{Rate of serpentinization in seafloor envrionments}.
\newblock \bibinfo{journal}{Tectonophysics} \bibinfo{volume}{116},
  \bibinfo{pages}{123--135, doi:10.1016/0040--1951(85)90225--2}.
 %Type = Article
\bibitem[{Mackenzie et~al.(2008)Mackenzie, Iess, Tortora and Rappaport}]{Mackenzie2008}
\bibinfo{author}{Mackenzie, R.A.}, \bibinfo{author}{Iess, L.},
  \bibinfo{author}{Tortora, P.}, \bibinfo{author}{Rappaport, N.J.}, \bibinfo{year}{2008}.
\newblock \bibinfo{title}{A non-hydrostatic {Rhea}}.
\newblock \bibinfo{journal}{Geophys. Res. Lett.} \bibinfo{volume}{35},
  \bibinfo{pages}{L05204, doi:10.1029/2007GL032898}.
%Type = Article
\bibitem[{McKinnon(1997)}]{McKinnon1997}
\bibinfo{author}{McKinnon, W.B.}, \bibinfo{year}{1997}.
\newblock \bibinfo{title}{{Mystery of Callisto: Is it undifferentiated?}}
\newblock \bibinfo{journal}{Icarus} \bibinfo{volume}{130},
  \bibinfo{pages}{540--543, doi:10.1006/icar.1997.5826}.
%Type = Article
\bibitem[{McKinnon and Bland(2011)}]{McKinnon2011}
\bibinfo{author}{McKinnon, W.B.}, \bibinfo{author}{Bland, M.T.},
  \bibinfo{year}{2011}.
\newblock \bibinfo{title}{{Core evolution in icy satellites and Kuiper belt
  objects}}.
\newblock \bibinfo{journal}{LPSC XLII Abstracts $\#$2768.}
%Type = Article
\bibitem[{Mitri et~al.(2010)Mitri, Pappalardo and Stevenson}]{Mitri2010}
\bibinfo{author}{Mitri, G.}, \bibinfo{author}{Pappalardo, R.T.},
  \bibinfo{author}{Stevenson, D.J.}, \bibinfo{year}{2010}.
\newblock \bibinfo{title}{{Evolution and interior structrure of Titan}}.
\newblock \bibinfo{journal}{LPSC XLI Abstracts $\#$2229}.
%Type = Article
\bibitem[{Mitri and Showman(2008)}]{Mitri2008}
\bibinfo{author}{Mitri, G.}, \bibinfo{author}{Showman, A.P.},
  \bibinfo{year}{2008}.
\newblock \bibinfo{title}{{Thermal convection in ice-I shells of Titan and
  Enceladus}}.
\newblock \bibinfo{journal}{Icarus} \bibinfo{volume}{193},
  \bibinfo{pages}{387--396, doi:10.1016/j.icarus.2007.07.016}.
%Type = Article
\bibitem[{Mosqueira and Estrada(2003a)}]{Mosqueira2003a}
\bibinfo{author}{Mosqueira, I.}, \bibinfo{author}{Estrada, P.R.},
  \bibinfo{year}{2003}a.
\newblock \bibinfo{title}{{Formation of the regular satellites of giant planets
  in an extended gaseous nebula I: subnebula model and accretion of
  satellites}}.
\newblock \bibinfo{journal}{Icarus} \bibinfo{volume}{163},
  \bibinfo{pages}{198--231, doi:10.1016/S0019--1035(03)00076--9}.
%Type = Article
\bibitem[{Mosqueira and Estrada(2003b)}]{Mosqueira2003b}
\bibinfo{author}{Mosqueira, I.}, \bibinfo{author}{Estrada, P.R.},
  \bibinfo{year}{2003}b.
\newblock \bibinfo{title}{{Formation of the regular satellites of giant planets
  in an extended gaseous nebula II: satellite migration and survival}}.
\newblock \bibinfo{journal}{Icarus} \bibinfo{volume}{163},
  \bibinfo{pages}{232--255, doi:10.1016/S0019--1035(03)00077--0}.
%Type = Article
\bibitem[{Mosqueira et~al.(2010)Mosqueira, Estrada and Charnoz}]{Mosqueira2010}
\bibinfo{author}{Mosqueira, I.}, \bibinfo{author}{Estrada, P.R.},
  \bibinfo{author}{Charnoz, S.}, \bibinfo{year}{2010}.
\newblock \bibinfo{title}{{Deciphering the origin of the regular satellites of
  gaseous giants --- Iapetus: The Rosetta ice-moon}}.
\newblock \bibinfo{journal}{Icarus} \bibinfo{volume}{207},
  \bibinfo{pages}{448--460, doi:10.1016/j.icarus.2009.10.018}.
%Type = Article
\bibitem[{Mueller and McKinnon(1988)}]{Mueller1988}
\bibinfo{author}{Mueller, S.}, \bibinfo{author}{McKinnon, W.B.},
  \bibinfo{year}{1988}.
\newblock \bibinfo{title}{Three-layered models of {Ganymede and Callisto:
  Compositions,} structures, and aspects of evolution}.
\newblock \bibinfo{journal}{Icarus} \bibinfo{volume}{76},
  \bibinfo{pages}{437--464}.
%Type = Article
\bibitem[{Nagel et~al.(2004)Nagel, Breuer and Spohn}]{Nagel2004}
\bibinfo{author}{Nagel, K.}, \bibinfo{author}{Breuer, D.},
  \bibinfo{author}{Spohn, T.}, \bibinfo{year}{2004}.
\newblock \bibinfo{title}{A model for the interior structure, evolution, and
  differentiation of {Callisto}}.
\newblock \bibinfo{journal}{Icarus} \bibinfo{volume}{169},
  \bibinfo{pages}{402--412, doi:10.1016/j.icarus.2003.12.019}.
%Type = Article
\bibitem[{Niemann et~al.(2005)Niemann, Atreya, Bauer, Carignan, Demick, Frost,
  Gautier, Haberman, Harpold, Hunten, Israel, Lunine, Kasprzak, Owen,
  Paulkovich, Raulin, Raaen and Way}]{Niemann2005}
\bibinfo{author}{Niemann, H.B.}, \bibinfo{author}{Atreya, S.K.},
  \bibinfo{author}{Bauer, S.J.}, \bibinfo{author}{Carignan, G.R.},
  \bibinfo{author}{Demick, J.E.}, \bibinfo{author}{Frost, R.L.},
  \bibinfo{author}{Gautier, D.}, \bibinfo{author}{Haberman, J.A.},
  \bibinfo{author}{Harpold, D.N.}, \bibinfo{author}{Hunten, D.M.},
  \bibinfo{author}{Israel, G.}, \bibinfo{author}{Lunine, J.I.},
  \bibinfo{author}{Kasprzak, W.T.}, \bibinfo{author}{Owen, T.C.},
  \bibinfo{author}{Paulkovich, M.}, \bibinfo{author}{Raulin, F.},
  \bibinfo{author}{Raaen, E.}, \bibinfo{author}{Way, S.H.},
  \bibinfo{year}{2005}.
\newblock \bibinfo{title}{{The abundances of constituents of Titan's atmospehre
  from the GCMS instrument on the Huygens probe}}.
\newblock \bibinfo{journal}{Nature} \bibinfo{volume}{438},
  \bibinfo{pages}{779--784, doi:10.1038/nature04122}.
%Type = Article
\bibitem[{Nimmo and Bills(2010)}]{Nimmo2010}
\bibinfo{author}{Nimmo, F.}, \bibinfo{author}{Bills, B.G.},
  \bibinfo{year}{2010}.
\newblock \bibinfo{title}{{Shell thickness variations and the long-wavelength
  topography of Titan}}.
\newblock \bibinfo{journal}{Icarus} \bibinfo{volume}{208},
  \bibinfo{pages}{896--904, doi:10.1016/j.icarus.2010.02.020}.
%Type = Article
\bibitem[{Noguchi and Niino(2010)}]{Noguchi2010}
\bibinfo{author}{Noguchi, T.}, \bibinfo{author}{Niino, H.},
  \bibinfo{year}{2010}.
\newblock \bibinfo{title}{{Multi-layered diffusive convection. Part 1.
  Spontaneous layer formation}}.
\newblock \bibinfo{journal}{J. Fluid Mech.} \bibinfo{volume}{651},
  \bibinfo{pages}{443--464, doi:10.1017/S0022112009994150}.
%Type = Article
\bibitem[{Perrillat et~al.(2005)Perrillat, Daniel, Koga, Reynard, Cardon and
  Crichton}]{Perrillat2005}
\bibinfo{author}{Perrillat, J.P.}, \bibinfo{author}{Daniel, I.},
  \bibinfo{author}{Koga, K.T.}, \bibinfo{author}{Reynard, B.},
  \bibinfo{author}{Cardon, H.}, \bibinfo{author}{Crichton, W.A.},
  \bibinfo{year}{2005}.
\newblock \bibinfo{title}{{Kinetics of antigorite dehydration: A real-time
  X-ray diffraction study}}.
\newblock \bibinfo{journal}{Earth Planet. Sci. Lett.} \bibinfo{volume}{236},
  \bibinfo{pages}{899--913, doi:10.1016/j.epsl.2005.06.006}.
%Type = Book
\bibitem[{Petrenko and Whitworth(1999)}]{Petrenko1999}
\bibinfo{author}{Petrenko, V.F.}, \bibinfo{author}{Whitworth, R.W.},
  \bibinfo{year}{1999}.
\newblock \bibinfo{title}{Physics of Ice}.
\newblock \bibinfo{publisher}{Oxford University Press}.
%Type = Article
\bibitem[{Ransford et~al.(1981)Ransford, Finnerty and Collerson}]{Ransford1981}
\bibinfo{author}{Ransford, G.A.}, \bibinfo{author}{Finnerty, A.A.},
  \bibinfo{author}{Collerson, K.D.}, \bibinfo{year}{1981}.
\newblock \bibinfo{title}{Europa's petrological thermal history}.
\newblock \bibinfo{journal}{Nature} \bibinfo{volume}{289},
  \bibinfo{pages}{21--24, doi:10.1038/289021a0}.
%Type = Incollection
\bibitem[{Raulin et~al.(2009)Raulin, McKay, Lunine and Owen}]{Raulin2009}
\bibinfo{author}{Raulin, F.}, \bibinfo{author}{McKay, C.},
  \bibinfo{author}{Lunine, J.}, \bibinfo{author}{Owen, T.C.},
  \bibinfo{year}{2009}.
\newblock \bibinfo{title}{Titan's astrobiology}, in: \bibinfo{editor}{Brown,
  R.H.}, \bibinfo{editor}{Lebreton, J.P.}, \bibinfo{editor}{Waite, J.H.}
  (Eds.), \bibinfo{booktitle}{{Titan from Cassini-Huygens}}.
  \bibinfo{publisher}{Springer}. chapter~\bibinfo{chapter}{9}.
%Type = Article
\bibitem[{Rempel et~al.(2001)Rempel, Wettlaufer and Worster}]{Rempel2001}
\bibinfo{author}{Rempel, A.W.}, \bibinfo{author}{Wettlaufer, J.S.},
  \bibinfo{author}{Worster, M.G.}, \bibinfo{year}{2001}.
\newblock \bibinfo{title}{Interfacial premelting and the thermomolecular force:
  {Thermodynamic buoyancy}}.
\newblock \bibinfo{journal}{Phys. Rev. Lett.} \bibinfo{volume}{87},
  \bibinfo{pages}{088501, doi:10.1103/PhysRevLett.87.088501}.
%Type = Article
\bibitem[{Schmitt(1983)}]{Schmitt1983}
\bibinfo{author}{Schmitt, R.W.}, \bibinfo{year}{1983}.
\newblock \bibinfo{title}{The characteristics of salt fingers in a variety of
  fluid systems, including stellar interiors, liquid metals, oceans, and
  magmas}.
\newblock \bibinfo{journal}{Phys. Fluids} \bibinfo{volume}{26},
  \bibinfo{pages}{2373--2377}.
%Type = Article
\bibitem[{Schmitt(1994)}]{Schmitt1994}
\bibinfo{author}{Schmitt, R.W.}, \bibinfo{year}{1994}.
\newblock \bibinfo{title}{Double diffusion in oceanography}.
\newblock \bibinfo{journal}{Annu. Rev. Fluid Mech.} \bibinfo{volume}{26},
  \bibinfo{pages}{255--285}.
%Type = Article
\bibitem[{Scott et~al.(2002)Scott, Williams and Ryerson}]{Scott2002}
\bibinfo{author}{Scott, H.P.}, \bibinfo{author}{Williams, Q.},
  \bibinfo{author}{Ryerson, F.J.}, \bibinfo{year}{2002}.
\newblock \bibinfo{title}{Experimental constraints on the chemical evolution of
  large icy satellites}.
\newblock \bibinfo{journal}{Earth Planet. Sci. Lett.} \bibinfo{volume}{203},
  \bibinfo{pages}{399--412}.
%Type = Article
\bibitem[{Seipold and Schilling(2003)}]{Seipold2003}
\bibinfo{author}{Seipold, U.}, \bibinfo{author}{Schilling, F.R.},
  \bibinfo{year}{2003}.
\newblock \bibinfo{title}{Heat transport in serpentinites}.
\newblock \bibinfo{journal}{Tectonophysics} \bibinfo{volume}{370},
  \bibinfo{pages}{147--162, doi:10.1016/S0040--1951(03)00183--5}.
%Type = Article
\bibitem[{Simoes et~al.(2012)Simoes, Pfaff, Hamelin, Klenzing, Freudenreich,
  Beghin, Berthelier, Bromund, Grard, Lebreton, Martin, Rowland, Sentman,
  Takahashi and Yair}]{Simoes2012}
\bibinfo{author}{Simoes, F.}, \bibinfo{author}{Pfaff, R.},
  \bibinfo{author}{Hamelin, M.}, \bibinfo{author}{Klenzing, J.},
  \bibinfo{author}{Freudenreich, H.}, \bibinfo{author}{Beghin, C.},
  \bibinfo{author}{Berthelier, J.J.}, \bibinfo{author}{Bromund, K.},
  \bibinfo{author}{Grard, R.}, \bibinfo{author}{Lebreton, J.P.},
  \bibinfo{author}{Martin, S.}, \bibinfo{author}{Rowland, D.},
  \bibinfo{author}{Sentman, D.}, \bibinfo{author}{Takahashi, Y.},
  \bibinfo{author}{Yair, Y.}, \bibinfo{year}{2012}.
\newblock \bibinfo{title}{{Using Schumann resonance measurements for
  constraining the water abundance on the giant planets---Implications for the
  Solar System's formation}}.
\newblock \bibinfo{journal}{Astrophys. J.} \bibinfo{volume}{750},
  \bibinfo{pages}{85, doi:10.1088/0004--637X/750/1/85}.
%Type = Article
\bibitem[{Sohl et~al.(2003)Sohl, Hussmann, Schwentker, Spohn and
  Lorenz}]{Sohl2003}
\bibinfo{author}{Sohl, F.}, \bibinfo{author}{Hussmann, H.},
  \bibinfo{author}{Schwentker, B.}, \bibinfo{author}{Spohn, T.},
  \bibinfo{author}{Lorenz, R.D.}, \bibinfo{year}{2003}.
\newblock \bibinfo{title}{{Interior structure models and tidal Love numbers of
  Titan}}.
\newblock \bibinfo{journal}{J. Geophys. Res.} \bibinfo{volume}{108},
  \bibinfo{pages}{5130, doi:10.1029/2003JE002044}.
%Type = Article
\bibitem[{Sohl et~al.(2002)Sohl, Spohn, Breuer and Nagel}]{Sohl2002}
\bibinfo{author}{Sohl, F.}, \bibinfo{author}{Spohn, T.},
  \bibinfo{author}{Breuer, D.}, \bibinfo{author}{Nagel, K.},
  \bibinfo{year}{2002}.
\newblock \bibinfo{title}{{Implications from Galileo observations on the
  interior structure and chemistry of the Galilean satellites}}.
\newblock \bibinfo{journal}{Icarus} \bibinfo{volume}{157},
  \bibinfo{pages}{104--119, doi:10.1006/icar.2002.6828}.
%Type = Incollection
\bibitem[{Sotin et~al.(2009)Sotin, Mitri, Rappaport, Schubert and
  Stevenson}]{Sotin2009}
\bibinfo{author}{Sotin, C.}, \bibinfo{author}{Mitri, G.},
  \bibinfo{author}{Rappaport, N.J.}, \bibinfo{author}{Schubert, G.},
  \bibinfo{author}{Stevenson, D.J.}, \bibinfo{year}{2009}.
\newblock \bibinfo{title}{Titan's interior structure}, in:
  \bibinfo{editor}{Brown, R.H.}, \bibinfo{editor}{Lebreton, J.P.},
  \bibinfo{editor}{Waite, J.H.} (Eds.), \bibinfo{booktitle}{{Titan from
  Cassini-Huygens}}. \bibinfo{publisher}{Springer}.
  chapter~\bibinfo{chapter}{4}.
%Type = Article
\bibitem[{Spiegel(1972)}]{Spiegel1972}
\bibinfo{author}{Spiegel, E.A.}, \bibinfo{year}{1972}.
\newblock \bibinfo{title}{Convection in stars {II.} {Special effects}}.
\newblock \bibinfo{journal}{Ann. Rev. Astron. Astophys.} \bibinfo{volume}{10},
  \bibinfo{pages}{261--3044}.
%Type = Article
\bibitem[{Stern(1960)}]{Stern1960}
\bibinfo{author}{Stern, M.E.}, \bibinfo{year}{1960}.
\newblock \bibinfo{title}{The ``salt-fountain" and thermohaline convection}.
\newblock \bibinfo{journal}{Tellus} \bibinfo{volume}{12},
  \bibinfo{pages}{172--175}.
%Type = Article
\bibitem[{Tobie et~al.(2012)Tobie, Gautier and Hersant}]{Tobie2012}
\bibinfo{author}{Tobie, G.}, \bibinfo{author}{Gautier, D.},
  \bibinfo{author}{Hersant, F.}, \bibinfo{year}{2012}.
\newblock \bibinfo{title}{{Titan's bulk composition constrained by
  \textit{Cassini-Huygens}: Implication for internal outgassing}}.
\newblock \bibinfo{journal}{Astrophys. J.} \bibinfo{volume}{752},
  \bibinfo{pages}{125, doi:10.1088/0004--637X/752/2/125}.
%Type = Article
\bibitem[{Tobie et~al.(2006)Tobie, Lunine and Sotin}]{Tobie2006}
\bibinfo{author}{Tobie, G.}, \bibinfo{author}{Lunine, J.I.},
  \bibinfo{author}{Sotin, C.}, \bibinfo{year}{2006}.
\newblock \bibinfo{title}{Episodic outgassing as the origin of atmospheric
  methane on {Titan}}.
\newblock \bibinfo{journal}{Nature} \bibinfo{volume}{440},
  \bibinfo{pages}{61--64, doi:10.1038/nature04497}.
%Type = Article
\bibitem[{Tobie et~al.(2005)Tobie, Mocquet and Sotin}]{Tobie2005}
\bibinfo{author}{Tobie, G.}, \bibinfo{author}{Mocquet, A.},
  \bibinfo{author}{Sotin, C.}, \bibinfo{year}{2005}.
\newblock \bibinfo{title}{{Tidal dissipation within large icy satellites:
  Applications to Europa and Titan}}.
\newblock \bibinfo{journal}{Icarus} \bibinfo{volume}{177},
  \bibinfo{pages}{534--549, doi:10.1016/j.icarus.2005.04.006}.
%Type = Article
\bibitem[{Travis and Schubert(2005)}]{Travis2005}
\bibinfo{author}{Travis, B.J.}, \bibinfo{author}{Schubert, G.},
  \bibinfo{year}{2005}.
\newblock \bibinfo{title}{Hydrothermal convection in carbonaceous chondrite
  parent bodies}.
\newblock \bibinfo{journal}{Earth Planet. Sci. Lett.} \bibinfo{volume}{240},
  \bibinfo{pages}{234--250, doi:10.1016/j.epsl.2005.09.008}.
%Type = Article
\bibitem[{Tsiganis et~al.(2005)Tsiganis, Gomes, Morbidelli and
  Levison}]{Tsiganis2005}
\bibinfo{author}{Tsiganis, K.}, \bibinfo{author}{Gomes, R.},
  \bibinfo{author}{Morbidelli, A.}, \bibinfo{author}{Levison, H.F.},
  \bibinfo{year}{2005}.
\newblock \bibinfo{title}{{Origin of the orbital architecture of the giant
  planets of the Solar System}}.
\newblock \bibinfo{journal}{Nature} \bibinfo{volume}{435},
  \bibinfo{pages}{459--461, doi:10.1038/nature03539}.
%Type = Book
\bibitem[{Turner(1973)}]{Turner1973}
\bibinfo{author}{Turner, J.S.}, \bibinfo{year}{1973}.
\newblock \bibinfo{title}{Buoyancy Effects in Fluids}.
\newblock \bibinfo{publisher}{Cambridge University Press},
  \bibinfo{address}{New York: New York}.
%Type = Article
\bibitem[{Turner(1974)}]{Turner1974}
\bibinfo{author}{Turner, J.S.}, \bibinfo{year}{1974}.
\newblock \bibinfo{title}{Double-diffusive phenomena}.
\newblock \bibinfo{journal}{Ann. Rev. Fluid Mech.} \bibinfo{volume}{6},
  \bibinfo{pages}{37--54}.
%Type = Article
\bibitem[{Turner(1985)}]{Turner1985}
\bibinfo{author}{Turner, J.S.}, \bibinfo{year}{1985}.
\newblock \bibinfo{title}{Multicomponent convection}.
\newblock \bibinfo{journal}{Ann. Rev. Fluid Mech.} \bibinfo{volume}{17},
  \bibinfo{pages}{11--44}.
%Type = Article
\bibitem[{Veronis(1968)}]{Veronis1968}
\bibinfo{author}{Veronis, G.}, \bibinfo{year}{1968}.
\newblock \bibinfo{title}{Effect of a stabilizing gradient of solute on thermal
  convection}.
\newblock \bibinfo{journal}{J. Fluid Mech.} \bibinfo{volume}{34},
  \bibinfo{pages}{315--336}.
%Type = Article
\bibitem[{Worster(2004)}]{Worster2004}
\bibinfo{author}{Worster, M.G.}, \bibinfo{year}{2004}.
\newblock \bibinfo{title}{Time-dependent fluxes across double-diffusive
  interfaces}.
\newblock \bibinfo{journal}{J. Fluid Mech.} \bibinfo{volume}{505},
  \bibinfo{pages}{287--307, doi:10.1017/S0022112004008523}.

\end{thebibliography}

%% not want to use model2-names.bst.



%% \bibitem must have one of the following forms:
%%   \bibitem[Jones et al.(1990)]{key}...
%%   \bibitem[Jones et al.(1990)Jones, Baker, and Williams]{key}...
%%   \bibitem[Jones et al., 1990]{key}...
%%   \bibitem[\protect\citeauthoryear{Jones, Baker, and Williams}{Jones
%%       et al.}{1990}]{key}...
%%   \bibitem[\protect\citeauthoryear{Jones et al.}{1990}]{key}...
%%   \bibitem[\protect\astroncite{Jones et al.}{1990}]{key}...
%%   \bibitem[\protect\citename{Jones et al., }1990]{key}...
%%   \harvarditem[Jones et al.]{Jones, Baker, and Williams}{1990}{key}...
%%


% \end{thebibliography}

\end{document}